\documentclass{aastex631}

\usepackage{mathtools}
\usepackage{amsmath}
\usepackage{physics}
\usepackage{booktabs}

\begin{document}

\title{Cosmic ray north-south anisotropy: rigidity spectrum and solar cycle variations observed by ground-based muon detectors}
\correspondingauthor{Masayoshi Kozai}
\email{kozai.masayoshi@nipr.ac.jp}

\author[0000-0002-3948-3666]{M. Kozai}
\affiliation{Polar Environment Data Science Center (PEDSC), Joint Support-Center for Data Science Research (ROIS-DS), Research Organization of Information and Systems, Tachikawa, Tokyo, Japan}

\author[0000-0002-0890-0607]{Y. Hayashi}
\affiliation{Physics Department, Shinshu University, Matsumoto, Nagano, Japan}

\author[0000-0002-4315-6369]{K. Fujii}
\affiliation{National Institute of Informatics (NII), Research Organization of Information and Systems, Chiyoda-ku, Tokyo, Japan}

\author[0000-0002-2131-4100]{K. Munakata}
\affiliation{Physics Department, Shinshu University, Matsumoto, Nagano, Japan}

\author[0000-0002-4913-8225]{C. Kato}
\affiliation{Physics Department, Shinshu University, Matsumoto, Nagano, Japan}

\author[0009-0006-3569-7380]{N. Miyashita}
\affiliation{Physics Department, Shinshu University, Matsumoto, Nagano, Japan}

\author[0000-0002-6105-9562]{A. Kadokura}
\affiliation{Polar Environment Data Science Center (PEDSC), Joint Support-Center for Data Science Research (ROIS-DS), Research Organization of Information and Systems, Tachikawa, Tokyo, Japan}

\author[0000-0001-9400-1765]{R. Kataoka}
\affiliation{National Institute of Polar Research (NIPR), Research Organization of Information and Systems, Tachikawa, Tokyo, Japan}

\author[0000-0002-3067-655X]{S. Miyake}
\affiliation{National Institute of Technology (KOSEN), Gifu College, Motosu, Gifu, Japan}

\author[0000-0001-7463-8267]{M.L. Duldig}
\affiliation{School of Natural Sciences, University of Tasmania, Hobart, Tasmania, Australia}

\author[0000-0002-4698-1671]{J.E. Humble}
\affiliation{School of Natural Sciences, University of Tasmania, Hobart, Tasmania, Australia}

\author[0000-0002-2464-5212]{K. Iwai}
\affiliation{Institute for Space-Earth Environmental Research (ISEE), Nagoya University, Nagoya, Aichi, Japan}

\begin{abstract}

The north-south (NS) anisotropy of galactic cosmic rays (GCRs) is dominated by a diamagnetic drift flow of GCRs in the interplanetary magnetic field (IMF), allowing us to derive key parameters of cosmic-ray propagation, such as the density gradient and diffusion coefficient.
We propose a new method to analyze the rigidity spectrum of GCR anisotropy and reveal a solar cycle variation of the NS anisotropy's spectrum using ground-based muon detectors in Nagoya, Japan, and Hobart, Australia.
The physics-based correction method for the atmospheric temperature effect on muons is used to combine the different-site detectors free from local atmospheric effects.
NS channel pairs in the multi-directional muon detectors are formed to enhance sensitivity to the NS anisotropy, and in this process, general graph matching in graph theory is introduced to survey optimized pairs.
Moreover, Bayesian estimation with the Gaussian process allows us to unfold the rigidity spectrum without supposing any analytical function for the spectral shape.
Thanks to these novel approaches, it has been discovered that the rigidity spectrum of the NS anisotropy is dynamically varying with solar activity every year.
It is attributed to a rigidity-dependent variation of the radial density gradient of GCRs based on the nature of the diamagnetic drift in the IMF.
The diffusion coefficient and mean-free-path length of GCRs as functions of the rigidity are also derived from the diffusion-convection flow balance.
This analysis expands the estimation limit of the mean-free-path length into $\le200$ GV rigidity region from $<10$ GV region achieved by solar energetic particle observations.

\end{abstract}


\section{Introduction} \label{sec:intro}
The cosmic-ray intensity from $\sim0.1$ to $\sim100$ GV rigidity region shows a clear anti-correlation with the solar activity typically represented by the sunspot number, explicitly indicating that galactic cosmic rays (GCRs) are influenced by the helispheric state, or solar modulation, through which they propagate before approaching Earth.
Measurements of GCR properties, including the intensity, energy spectrum, composition, anisotropy, and their temporal variations, on the ground and in the space have been driving the elucidation of the GCR propagation and space environment.
The anisotropy, or momentum-space distribution of GCRs, is expected to be the next key to reveal solar modulation;
although it is theoretically proved to have a close relation with the interplanetary magnetic field (IMF) structure and GCR diffusion coefficients \citep{gleeson1969,munakata1986}, observed solar cycle variations of the anisotropy are nearly unexplored in quantitative reproductions by heliospheric simulations, unlike the GCR density variations.
Even in the broader areas of cosmic-ray physics, including galactic propagation, a consistent interpretation of the anisotropy with other properties has been an open question \citep{gabici2019,gabici2023}.

Ground-based observations of cosmic rays measure secondary cosmic rays produced by GCR interactions with atmospheric nuclei and the subsequent reactions.
They feature long-term stability, improved statistics from large detection areas, and large angular acceptances by a worldwide network, all essential to study solar modulation phenomena consisting of diverse temporal variations from hourly (e.g., interplanetary shocks) to $>10$-year (solar cycles) scales.
Among secondary cosmic-ray species on the ground, muons have the highest flux (statistics) and relatively accurate response of the incident angle to primary cosmic rays because they directly originate from pion decays immediately occurring after the atmospheric nuclear spallations by primary cosmic rays typically around the upper edge of troposphere.
These advantages make the ground-based muon detector a unique means for measuring the anisotropy whose magnitude is only $\sim0.1$\% of GCR intensity.

However, anisotropy observation by muon detectors has suffered from an atmospheric temperature variation that perturbs muon counting rates.
Conventional methods \citep{mori1979,fujimoto1984,munakata1997,okazaki2008,munakata2014,kozai2016} have corrected the temperature effect by subtracting muon counting rates of directional channels from each other to cancel out the effect.
The subtracted directional channels must be from the identical location where the temperature effect is expected to be similar.
Additionally, in some analyses, the statistics are degraded by this subtraction.
On the other hand, \cite{mendonca2016,mendonca2019} proved the validity of a more physics-based approach in which the temperature effect is directly corrected in each directional channel by using meteorological reanalysis data.
Contrary to the conventional method, this correction method allows for a direct combination of observations from different sites free from the local temperature effect.
\cite{munakata2022} applied this correction to derive the rigidity spectrum of the anisotropy in a solar eruption event, demonstrating the new temperature correction method for anisotropy studies.
However, there is still a problem in that power-law indices of the rigidity spectra of the derived anisotropy have unreasonably unstable fluctuations, especially in the case of a small anisotropy amplitude.
The power-law function supposed to reconstruct the rigidity spectrum in their study is likely too definitive for the dynamically varying spectrum, causing this problem.

In this study, we propose a new method applying Bayesian estimation for deriving the rigidity spectrum.
It does not pre-suppose an analytical function such as the power-law function for the spectrum, while it uses the Gaussian process to confine the smoothness of the spectrum instead.
This high tolerance of varying spectral shapes enables us to trace the dynamic variation of the spectrum in solar modulation phenomena and provides a reliable analysis result.
Additionally, we introduce general graph matching \citep{edmonds1965} in graph theory to survey optimal channel combinations in the inter-station network, leveraging the advantage of the temperature correction method that allows us to combine multiple station data directly.
We demonstrate these analysis ideas by analyzing the rigidity spectrum of the north-south (NS) anisotropy observed by ground-based muon detectors in Nagoya, Japan, and Hobart, Australia.
The NS anisotropy, or NS flow of GCRs, shows a polarity reversal according to the local IMF sector assigned as $Toward$ or $Away$ each time \citep{swinson1969,swinson1971,yasue1980,chen1993,hall1994,laurenza2003,kozai2014,munakata2014}.
This phenomenon proves that the NS anisotropy is dominated by a diamagnetic drift flow of GCRs induced by a combination of the gyro-motion and radial (anti-sunward) density gradient of GCRs in the IMF.
Based on this mechanism, we derive rigidity-dependent variations of the density gradient and, subsequently, the radial diffusion coefficient of GCRs, which are essential parameters to elucidate solar modulation.
We also report a plan for the NS conjugate observation in the polar region, which will vastly improve the sensitivity of the muon detector network to the NS anisotropy.

\section{Method} \label{sec:method}

\subsection{Preprocessing of muon counting rates} \label{subsec:eta}
Table \ref{tab:detector} lists characteristics of the Nagoya and Hobart muon detectors used in this paper, compiled by \cite{munakata2022}.
It summarizes the geographical latitude, longitude, and altitude of the observation sites ($\lambda_D$, $\phi_D$, and alt.), the number of directional channels (ch-no.) each counting incident muons, the geomagnetic cutoff rigidity $P_{\rm cut}$, the hourly muon counting rate (cph), its statistical error $\sigma$ (0.01\%), the median rigidity $P_{\rm median}$ of primary cosmic rays producing muons measured in each directional channel, and the asymptotic trajectory direction ($\lambda_{\rm asymp}, \phi_{\rm asymp}$) of the primary cosmic rays with the median rigidity outside the magnetosphere.
Nagoya and Hobart stations started their observations in 1970 and 1992, respectively, and currently form a part of the Global Muon Detector Network (GMDN).
We analyze their muon counting rates from 1995 when both the muon detectors and the solar wind observation data used in this paper have sufficient duty cycles.
Archive data of the muon counting rates in GMDN are published in Shinshu University's institutional repository \citep{gmdncollaboration2024}.
\begin{table}
\centering
\caption{Characteristics of muon detectors}
\scriptsize
\begin{tabular}{lrrrrcccccc}
\toprule
Station & $\lambda_D$ (deg) & $\phi_D$ (deg) & alt.(m) & ch-no. & $P_{\rm cut}$ (GV) & cph/$10^4$ & $\sigma$ (0.01\%) & $P_{\rm median}$ (GV) & $\lambda_{\rm asymp}$ (deg) & $\phi_{\rm asymp}$ (deg)\\
\midrule
Nagoya & 35.2N & 137.0E & 77 & 17 & 8.0–12.6 & 17.3–285.6 & 5.9–24.0 & 58.4–106.9 & 64.0N–24.4S & 89.1E–235.0E \\
Hobart & 43.0S & 147.3E & 65 & 13 & 2.5–4.0 & 19.9–149.3 & 8.2–22.4 & 53.1–74.0 & 5.0N–76.6S & 122.4E–237.0E \\
\bottomrule
\end{tabular}
\label{tab:detector}
\end{table}

Our processing of muon counting rates described below partially refers to the conventional analysis method of the NS anisotropy, known as Nagoya-GG \citep{mori1979}.
It uses muon counting rates $I_i$ uncorrected for the temperature effect in Nagoya station and takes differences
\begin{equation}
    {\rm GG} = \frac{ (I_{\rm N2} - I_{\rm S2}) + (I_{\rm N2} - I_{\rm E2}) }{2}
    \label{eq:nagoya-gg}
\end{equation}
where factor 2 is introduced in this paper for a quantitative comparison with our method.
The channel identifiers N2, S2, and E2 indicate secondary-inclined channels whose central viewing directions are north-, south-, and eastward, respectively, with a zenith angle of 49$^\circ$.
The first NS differential term, $I_{\rm N2} - I_{\rm S2}$ ensures a high response of Nagoya-GG to the NS anisotropy.
The second term $I_{\rm N2} - I_{\rm E2}$ is introduced to cancel out the temperature effect remaining in the first term.
Nagoya-GG is proven to have sufficient sensitivity to the NS anisotropy and has contributed to revealing it \citep{yasue1980,laurenza2003,munakata2014,kozai2014}.

In our method, on the other hand, the atmospheric temperature effect on muons is corrected in each directional channel of Nagoya and Hobart detectors by a physics-based method using the meteorological reanalysis data, as described in Appendix \ref{sec:prepare}.
The corrected muon counting rate $I_i(t)$ for each directional channel $i$ is coupled to the GCR anisotropy in space as \citep{nagashima1971,fujimoto1984,kozai2016a}
\begin{equation}
    \begin{split}
    	I_i(t) \sim \int_{P=0}^\infty \bigl[ &\xi_0(t,P) \dd{c_{0,i}^0}(P) + \xi_z(t,P) \dd{c_{1,i}^0}(P) +\\
            &\xi_x(t,P) \qty{ \dd{c_{1,i}^1}(P) \cos \omega t_{\rm st} - \dd{s_{1,i}^1}(P) \sin \omega t_{\rm st} }
            + \xi_y(t,P) \qty{ \dd{s_{1,i}^1}(P) \cos \omega t_{\rm st} + \dd{c_{1,i}^1}(P) \sin \omega t_{\rm st} } \bigr] .
    \label{eq:I-xi}
    \end{split}
\end{equation}
Here, the anisotropy is expanded into the zeroth and first-order spherical harmonics in momentum space with rigidity $P$.
The expansion coefficients $\xi_0$, $\xi_z$, $\xi_x$ and $\xi_y$ are defined in a geocentric (GEO) coordinate system whose $z$-axis points geographic north pole and $x$-axis points to midnight in the equatorial plane.
Indices $n$ and $m$ attached to coefficients $\dd{c_{n,i}^m}$ and $\dd{s_{n,i}^m}$ are degree and order of the spherical harmonics, respectively, representing components of the zeroth and first-order harmonics.
These coefficients are known as the differential coupling coefficients and are derived from numerical calculations of GCR propagation in the magnetosphere, atmosphere, and detector \citep{kozai2016a}.
The local time $t_{\rm st}$ of each station to which the directional channel $i$ belongs is related to universal time $t$ as $\omega t_{\rm st} = \omega t + \phi_{\rm st}$ where $\phi_{\rm st}$ is the station's geographic longitude and $\omega=\pi/12$.
The higher order ($\ge2$nd) harmonics are omitted because they have negligible amplitudes compared to the zeroth and first-order anisotropy in the usual state.
This equation describes the coupling between the observed muon counting rate $I_i$ and the space anisotropy ($\xi_0$, $\xi_z$, $\xi_x$ and $\xi_y$) via the differential coupling coefficients at each rigidity $P$.
The zeroth order anisotropy $\xi_0$, or isotropic component, represents a variation of the GCR density from its Bartels rotation average because the counting rate $I_i$ is converted into a deviation from its Bartels rotation average as shown in Appendix \ref{sec:prepare}.
Equatorial components of the first-order anisotropy ($\xi_x,\xi_y$) represent a diurnal anisotropy or the GCR flow in the equatorial plane.
The remaining component $\xi_z$ is the NS anisotropy representing the NS component of the GCR flow, which is the main target of this study.

The hourly counting rate $I_i(t)$ is averaged for each day in which all counting rates and hourly IMF data mentioned below are available for at least 20 hours.
This daily average
$\boldsymbol{I_i^{\rm day}(t_{\rm day})}$
is expected to minimize a contribution from the diurnal anisotropy ($\xi_x, \xi_y$) in the counting rates.
The daily counting rates are sorted into days assigned to each IMF sector, $Toward$ or $Away$, and averaged over each sector's period in each year, as $I_i^T$ for the $Toward$ sector or $I_i^{A}$ for the $Away$ sector.
In this process, only Bartels rotations each with $\ge5$ $Toward$, $\ge5$ $Away$, and total $\ge15$ available days are used to secure the sector reversal in each Baterls rotation.
A standard deviation for the yearly counting rate ($I_i^T$ or $I_i^A$) is derived from the variance of $I_i^{\rm day}(t_{\rm day})$ in each sector's period in each year, and then the standard error, $\sigma_i^T$ or $\sigma_i^A$, is derived by dividing the standard deviation by a square root of the number of days assigned to each sector in each year.

The IMF data is obtained from OMNIWeb service \citep{papitashvili2020,king2005} in the geocentric solar ecliptic (GSE) coordinate system whose $x$-axis points toward Sun and $y$-axis opposes Earth's orbital motion.
In the picture of the diamagnetic drift dominating the NS anisotropy \citep{swinson1969}, the IMF parallel to the ecliptic plane is assumed, and its GSE-$y$ component ($B_y$) is essential for determining the anisotropy.
Therefore, we omit days with $|B_z|>|B_y|$ for the daily average IMF, and the IMF sector is identified daily as a $Toward$ ($Away$) sector when $B_y < 0$ ($B_y \ge 0$).
It is also noted that a definition of each year is slightly modified as described in Table \ref{tab:year}, so that its start and end dates correspond to boundaries of the Bartels rotation periods.
\begin{table}
\centering
\caption{Definition of each year in this paper}
\begin{tabular}{rlll}
\toprule
Year & Bartels rotations & Start date & End date \\
\midrule
1995 & 2205-2218 & Jan 12, 1995 & Jan 24, 1996 \\
1996 & 2219-2231 & Jan 25, 1996 & Jan 09, 1997 \\
1997 & 2232-2245 & Jan 10, 1997 & Jan 22, 1998 \\
1998 & 2246-2258 & Jan 23, 1998 & Jan 08, 1999 \\
1999 & 2259-2272 & Jan 09, 1999 & Jan 21, 2000 \\
2000 & 2273-2285 & Jan 22, 2000 & Jan 06, 2001 \\
2001 & 2286-2299 & Jan 07, 2001 & Jan 19, 2002 \\
2002 & 2300-2312 & Jan 20, 2002 & Jan 05, 2003 \\
2003 & 2313-2326 & Jan 06, 2003 & Jan 18, 2004 \\
2004 & 2327-2339 & Jan 19, 2004 & Jan 03, 2005 \\
2005 & 2340-2353 & Jan 04, 2005 & Jan 16, 2006 \\
2006 & 2354-2366 & Jan 17, 2006 & Jan 02, 2007 \\
2007 & 2367-2380 & Jan 03, 2007 & Jan 15, 2008 \\
2008 & 2381-2393 & Jan 16, 2008 & Dec 31, 2008 \\
2009 & 2394-2407 & Jan 01, 2009 & Jan 13, 2010 \\
2010 & 2408-2421 & Jan 14, 2010 & Jan 26, 2011 \\
2011 & 2422-2434 & Jan 27, 2011 & Jan 12, 2012 \\
2012 & 2435-2448 & Jan 13, 2012 & Jan 24, 2013 \\
2013 & 2449-2461 & Jan 25, 2013 & Jan 10, 2014 \\
2014 & 2462-2475 & Jan 11, 2014 & Jan 23, 2015 \\
2015 & 2476-2488 & Jan 24, 2015 & Jan 09, 2016 \\
2016 & 2489-2502 & Jan 10, 2016 & Jan 21, 2017 \\
2017 & 2503-2515 & Jan 22, 2017 & Jan 07, 2018 \\
2018 & 2516-2529 & Jan 08, 2018 & Jan 20, 2019 \\
2019 & 2530-2542 & Jan 21, 2019 & Jan 06, 2020 \\
2020 & 2543-2556 & Jan 07, 2020 & Jan 18, 2021 \\
2021 & 2557-2569 & Jan 19, 2021 & Jan 04, 2022 \\
2022 & 2570-2582 & Jan 05, 2022 & Dec 21, 2022 \\
\bottomrule
\end{tabular}
\label{tab:year}
\end{table}

Expanding the concept of Nagoya-GG into our temperature-corrected dataset, we take a difference between counting rates of directional channels $i$ and $j$ averaged for $Toward$ ($Away$) sector in each year as
\begin{equation}
    \eta_{ij}^{T(A)} = I_i^{T(A)} - I_j^{T(A)}.
\label{eq:eta-sec}
\end{equation}
Its error is derived from the error propagation, as
\begin{equation}
 \qty(\sigma_{ij}^{T(A)})^2 = \qty(\sigma_i^{T(A)})^2 + \qty(\sigma_j^{T(A)})^2 .
 \label{eq:sigma-eta-sec}
\end{equation}
The channel pair $ij$ is defined to have sufficient sensitivity to the NS anisotropy in Section \ref{subsec:pair}.
The sector reversal of the NS anisotropy is extracted by taking a difference between $\eta_{ij}^T$ and $\eta_{ij}^A$, and from equation (\ref{eq:I-xi}), it is expressed as
\begin{equation}
	\eta_{ij}^{TA} = \frac{\eta_{ij}^T - \eta_{ij}^A}{2} \sim \int_{P=0}^\infty
 \qty( \xi_z^{TA} \dd{c_{ij}^z} + \epsilon_0^{TA} \dd{c_{ij}^0} + \epsilon_c^{TA} \dd{c_{ij}^d} + \epsilon_s^{TA} \dd{s_{ij}^d} )
 \label{eq:eta-int}
\end{equation}
where
\begin{equation}
    \dd{c_{ij}^z} = \dd{c_{1,i}^0} - \dd{c_{1,j}^0},\quad
    \dd{c_{ij}^0} = \dd{c_{0,i}^0} - \dd{c_{0,j}^0},\quad
    \dd{c_{ij}^d} = \dd{c_{1,i}^1} - \dd{c_{1,j}^1},\qq{and}
    \dd{s_{ij}^d} = \dd{s_{1,i}^1} - \dd{s_{1,j}^1}.
\end{equation}
Rigidity $P$ as a variable of $\xi$ and $\dd{c}$ ($\dd{s}$) is omitted in this equation for simplicity.
A standard error of $\eta_{ij}^{TA}$ is derived as
\begin{equation}
    \qty(\sigma_{ij}^{TA})^2 = \frac{\qty(\sigma_{ij}^T)^2 + \qty(\sigma_{ij}^A)^2}{4} .
    \label{eq:sigma-eta}
\end{equation}
The parameter $\xi_z^{TA}$ represents a sector reversal of the NS anisotropy $\xi_z$, defined as $\xi_z^{TA} = \qty(\xi_z^T - \xi_z^A)/2$ where $\xi_z^{T(A)}$ denotes the $Toward$ ($Away$) sector average of $\xi_z$ in each year.
Hereafter, we call $\xi_z^{TA}(P)$ the rigidity spectrum of the NS anisotropy.
The parameters $\epsilon_0^{TA}$ and ($\epsilon_c^{TA},\epsilon_s^{TA}$) also correspond to $Toward$ - $Away$ sector differences of GCR density $\xi_0$ and diurnal anisotropy ($\xi_x,\xi_y$) respectively in each year.
Contrary to the NS anisotropy, these parameters have no physical reason to systematically depend on the IMF polarity in sufficiently long period averages.
Therefore these are expected to play only as perturbations on $\eta_{ij}^{TA}$, while $\xi_z^{TA}$ has a substantial contribution on $\eta_{ij}^{TA}$, leading to an approximation of equation (\ref{eq:eta-int}) as
\begin{equation}
    \eta_{ij}^{TA} \sim \int_{P=0}^\infty \xi_z^{TA} \dd{c_{ij}^z}.
    \label{eq:eta-xi}
\end{equation}
This equation indicates that the observed value $\eta_{ij}^{TA}$ is a convolution of the NS anisotropy spectrum $\xi_z^{TA}(P)$ through its differential response $\dd{c_{ij}^z(P)}/dP$.

Unlike Nagoya-GG described above, the channel pair $ij$ can be formed from any two of all channels in Nagoya and Hobart muon detectors, thanks to the physics-based temperature correction in each channel.
Nagoya and Hobart muon detectors have 17 and 13 directional channels, respectively;
the total 30 channels contain $30\times29/2=435$ kinds of channel pairs, and we need to select some pairs from them to analyze the NS anisotropy.
Simultaneously forming multiple pairs makes this problem more complex.
In the next section, we solve this problem using graph theory algorithm.

\subsection{Optimization of channel pairing} \label{subsec:pair}
Our optimization problem is defined as searching for a pattern of the channel pairing ($ij$'s) which maximizes the sum of sensitivity
\begin{equation}
    A = \sum_{ij} a_{ij}
\end{equation}
where $a_{ij}$ is a sensitivity of each channel pair $ij$ to the NS anisotropy, derived in Appendix \ref{sec:sense}.
The set of the pairs has to meet a condition that each channel ($i$ or $j$) does not duplicate in the set to ensure that derived $\eta_{ij}^{TA}$'s are independent of each other.
This problem is equivalent to the maximum matching problem in general graph, or general graph matching in graph theory, and solved by the \texttt{max\_weight\_matching} function of the Python \texttt{NetworkX} library \citep{networkx,galil1986,edmonds1965}.
\begin{figure}
    \centering
    \includegraphics[width=0.65\textwidth]{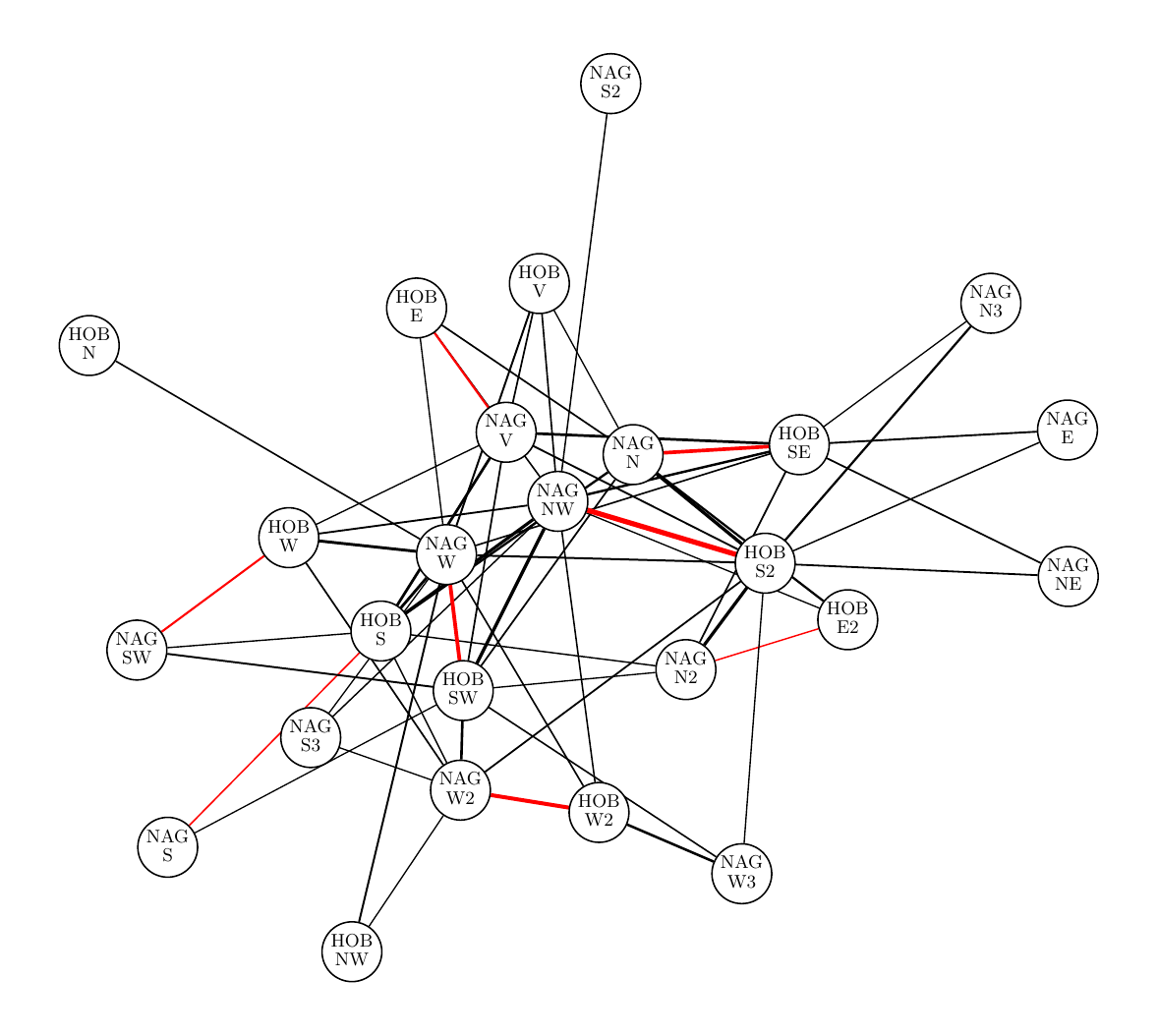}
    \caption{
    Visualization of general graph matching for optimizing channel pairs.
    Each node corresponds to a directional channel, and each edge connecting the nodes represents each channel pair.
    The line width of each edge is proportional to the sensitivity $a_{ij}$ of each pair $ij$, and red edges represent the pairs derived as the optimal solution.
    }
    \label{fig:network}
\end{figure}
Figure \ref{fig:network} visualizes this problem and its solution.
Each node in this network corresponds to a directional channel ($i$ or $j$) of each pair passing the threshold described in equation (\ref{eq:sense}), and each edge connecting the nodes represents each channel pair.
The node positions are determined by the Fruchterman-Reingold algorithm \citep{brodersen2023,fruchterman1991}.
The line width of each edge is proportional to the sensitivity $a_{ij}$ of each pair, and the edges with red colors represent the optimized pairing pattern which maximizes the total sensitivity $A$.

Each node's code ``NAG'' or ``HOB'' represents Nagoya or Hobart station.
Each character string below the station code indicates each directional channel, where V, N, S, E, and W indicate the vertical channel and north-, south-, east-, and westward inclined channels, respectively.
The 1st inclined channels (N, S, E, W) have central viewing direction with a zenith angle of 30$^\circ$, while the 2nd and 3rd inclined channels denoted by 2 and 3 in the channel name have the zenith angles of 49$^\circ$ and 60$^\circ$ respectively.
Table \ref{tab:pair} lists the derived channel pairs, $ij$'s, along with their coupling coefficient $c_{ij}^z$ for a flat spectrum and sensitivity $a_{ij}$ each derived as described in Appendix \ref{sec:sense}.
All left side channels $i$'s are those in Nagoya station in the northern hemisphere, while the other side channel $j$ subtracted from the channel $i$ in equation (\ref{eq:eta-sec}) is all in Hobart station in the southern hemisphere.
This is reasonable as a result of the optimization maximizing the NS anisotropy sensitivity.
\begin{table}
\centering
\caption{Channel pairs optimized for the NS anisotropy}
\begin{tabular}{rllrr}
\toprule
$l$ & Channel $i$ & Channel $j$ & Coupling coef. $c_{ij}^z$ & Sensitivity $a_{ij}$ \\
\midrule
1 & Nagoya-N2 & Hobart-E2 & 0.85 & 1.79 \\
2 & Nagoya-S & Hobart-S & 0.75 & 1.91 \\
3 & Nagoya-SW & Hobart-W & 0.74 & 2.58 \\
4 & Nagoya-V & Hobart-E & 0.72 & 2.71 \\
5 & Nagoya-N & Hobart-SE & 1.06 & 4.58 \\
6 & Nagoya-W & Hobart-SW & 1.16 & 4.72 \\
7 & Nagoya-W2 & Hobart-W2 & 0.98 & 4.97 \\
8 & Nagoya-NW & Hobart-S2 & 1.37 & 6.48 \\
\bottomrule
\end{tabular}
\label{tab:pair}
\end{table}

\begin{figure}
    \centering
    \includegraphics[width=1\textwidth]{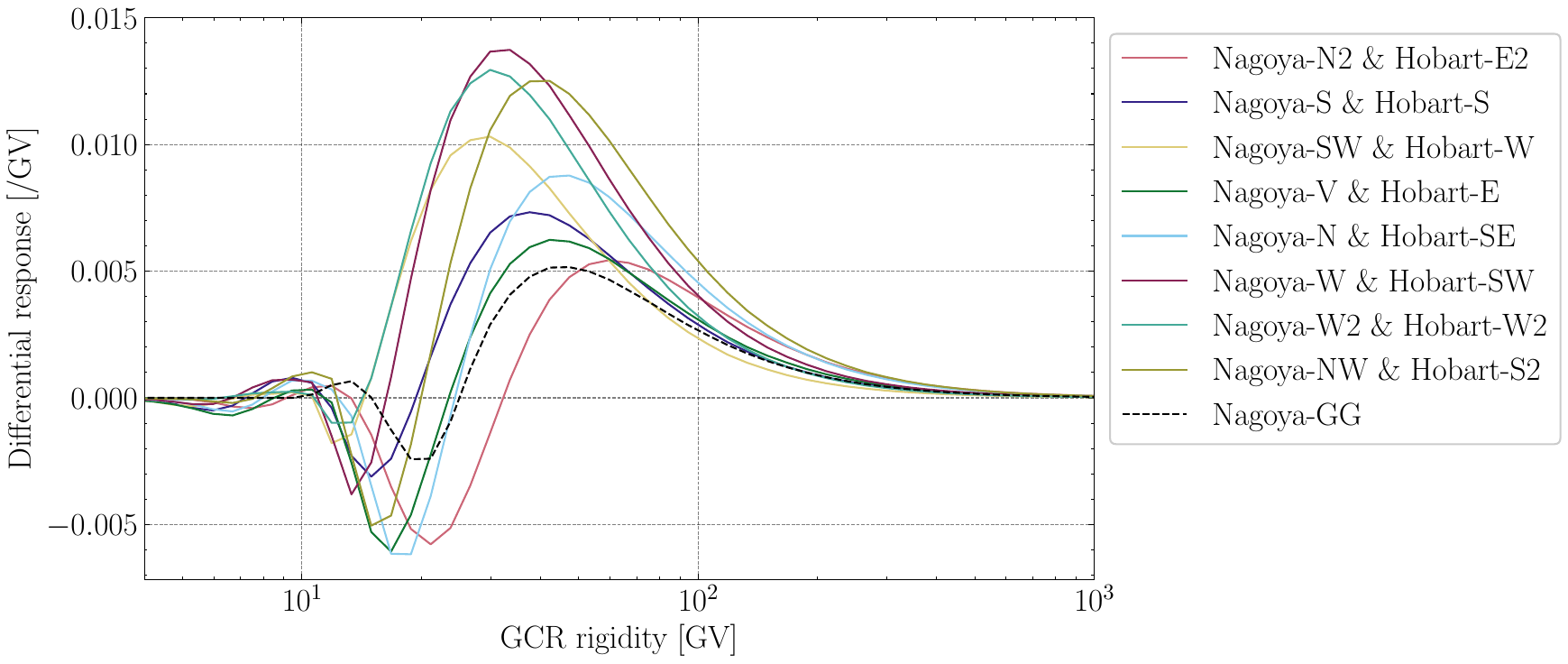}
    \caption{Differential response of each channel pair to the NS anisotropy.}
    \label{fig:dcpl}
\end{figure}
A solid line in Figure \ref{fig:dcpl} displays a differential response $\dd{c_{ij}^z}/\dd{P}$ for each channel pair $ij$ in table \ref{tab:pair}.
This differential response function has a positive value when the asymptotic direction of channel $i$ at rigidity $P$ points to a higher latitude (more northward) direction than channel $j$.
In other words, its negative value in the lower rigidity region in Figure \ref{fig:dcpl} indicates that the NS relation between channels $i$ and $j$ is swapped for low rigidity GCRs.
The low-rigidity GCRs passing through the mid- or low-latitude magnetosphere, such as those observed by Nagoya or Hobart detectors, are largely deflected from the original viewing angle of each directional channel by the geomagnetic field, causing this negative response.
This problem is solved by placing detectors in Arctic and Antarctic regions where the geomagnetic effect is minimized, and such an observation plan is mentioned in Section \ref{subsec:polar}.
The difference of the differential response between channel pairs displayed in Figure \ref{fig:dcpl} allows us to derive the rigidity spectrum $\xi_z^{TA}(P)$ of the NS anisotropy, as described in the next section.

Finally, the dashed line in Figure \ref{fig:dcpl} displays the differential response of Nagoya-GG for comparison.
The channel pairs $ij$'s used in this paper (solid lines) generally have higher response peaks than Nagoya-GG, thanks to the optimization of the channel pairing.

\subsection{Bayesian estimation of the rigidity spectrum} \label{subsec:bayes}
In conventional approaches, certain analytical functions are proposed for the rigidity spectrum of the anisotropy.
Most typically use a power-law function,
\begin{equation}
    \xi_z^{TA}(P) = \beta P^{\gamma},
    \label{eq:xi-gamma}
\end{equation}
and searches for parameters ($\beta, \gamma$) fitting to the observed counting rates \citep{yasue1980,hall1994,munakata2022}.
Although the power-law function is inapplicable for a spectrum crossing zero or with an insignificant magnitude, such a case can happen in the anisotropy component because the anisotropy is a vector quantity, unlike the absolute density spectrum.
\cite{munakata2022} suggests that this is a cause of the unstable fluctuations of the power-law index $\gamma$ in their results.
A single analytical function such as the power-law spectrum is probably too restrictive a premise to express all spectral shapes appearing in solar modulation phenomena. 
On the other hand, this paper derives the NS anisotropy spectrum and its error range by the Bayesian approach as described below.
The Gaussian process is introduced as a prior probability distribution in Bayesian estimation, confining acceptable ranges of the smoothness and magnitude of the spectrum instead of the analytical function.
This approach makes the spectrum estimation tolerant of varying spectral shapes, allowing us to trace dynamic variations of the spectrum.

Hereafter, the GCR rigidity $P$ [GV] is expressed by its logarithm $q = \log_{10}(P)$ in equations.
We approximate the rigidity spectrum by a step function with a finite number ($N$) of rigidity bins as
\begin{equation}
    \xi_z^{TA}(q) = \theta_k \qq{for} q_{k-1} \le q < q_k
    \label{eq:step}
\end{equation}
where $k=1,\ldots,N$ indicates each rigidity bin and $q_k = \log_{10} P_k$ represents an upper limit of the $k$-th rigidity bin.
The NS anisotropy $\theta_k$ in each rigidity bin $k$ is approximated to be constant for the interval $q_{k-1} \le q < q_k$.
The rigidity spectrum $\xi_z^{TA}(P)$ is now expressed by an $N$-dimension vector parameter $\boldsymbol{\theta} = \qty( \theta_1,\theta_2,\ldots,\theta_N)^\top$, and its mean and error can be expressed by a probability distribution in the $N$-dimension space in which each dimension corresponds to each rigidity bin $k$.
Based on Figure \ref{fig:dcpl}, our responsive rigidity range is from $\sim10$ to $\sim400$ GV, and we split this range into $N=8$ bins.
Table \ref{tab:bins} lists each rigidity bin with its index $k$ and upper boundary $q_k$.
Median rigidity $P_k^m$ [GV] of each rigidity bin satisfies $\log_{10} P_k^m = (q_{k-1} + q_k)/2$.
\begin{table}
\centering
\caption{Index $k$, upper rigidity limit $q_k$ in log-scale, upper rigidity limit $P_k$, and median rigidity $P_k^m$ for each rigidity bin. The $k=0$ row denotes only the lower rigidity limit of $k=1$ bin.}
\begin{tabular}{rrrr}
\toprule
$k$ & $q_k$ [$\log_{10}$(GV)] & $P_k$ [GV] & $P_k^m$ [GV] \\
\midrule
0 & 1.00 & 10.00 & NA \\
1 & 1.20 & 15.85 & 12.59 \\
2 & 1.40 & 25.12 & 19.95 \\
3 & 1.60 & 39.81 & 31.62 \\
4 & 1.80 & 63.10 & 50.12 \\
5 & 2.00 & 100.00 & 79.43 \\
6 & 2.20 & 158.49 & 125.89 \\
7 & 2.40 & 251.19 & 199.53 \\
8 & 2.60 & 398.11 & 316.23 \\
\bottomrule
\end{tabular}
\label{tab:bins}
\end{table}

The differential response $\dd{c_{ij}^z}/\dd{P}$ is also discretized and expressed by a matrix
\begin{equation}
    C_{lk} = \int_{P=P_{k-1}}^{P_k} \dd{c_{ij}^z(P)}
\end{equation}
where each channel pair $ij$ is replaced with $l = 1,2,\ldots,8$ in the index of Table \ref{tab:pair}.
From equation (\ref{eq:eta-xi}), expected value of the observable $\eta_l^{TA} = \eta_{ij}^{TA}$ from the parameter $\boldsymbol{\theta}$ is derived as
\begin{equation}
    \tilde\eta_l^{TA} = \sum_{k=1}^N C_{lk} \theta_k = \qty(\boldsymbol{C} \boldsymbol{\theta})_l.
    \label{eq:exp}
\end{equation}
Therefore, the conditional probability of the observed values $\eta_l^{TA}$'s for the parameter $\boldsymbol{\theta}$ is modeled as
\begin{equation}
    \begin{split}
        \mathcal{P}\qty(\boldsymbol{\eta} | \boldsymbol{\theta})
        &= \prod_l \mathcal{N}\qty(\eta_l^{TA} \;\middle|\; \tilde\eta_l^{TA}, \qty(\sigma_l^{TA})^2)\\
        &\propto \prod_l
        \exp[- \frac{\qty{\eta_l^{TA} - \qty(\boldsymbol{C} \boldsymbol{\theta})_l}^2}{2 \qty(\sigma_l^{TA})^2}]
        = \exp[-\frac{1}{2} \sum_l \frac{\qty{\eta_l^{TA} - \qty(\boldsymbol{C} \boldsymbol{\theta})_l}^2}{\qty(\sigma_l^{TA})^2}]
    \end{split}
    \label{eq:likelihood}
\end{equation}
where $\boldsymbol{\eta}$ is a vector consisting of $\eta_l^{TA}$'s for all channel pairs and $\sigma_l^{TA}=\sigma_{ij}^{TA}$ in equation (\ref{eq:sigma-eta}).
The symbol $\mathcal{N}(x\;|\;\mu,\sigma^2)$ expresses a normal distribution for a variable $x$ with its mean $\mu$ and variance $\sigma^2$.

On the other hand, the Gaussian process is introduced by considering the probability distribution of the spectrum $\xi_z^{TA}(q)$ as a multivariate normal distribution in which each dimension corresponds to each value of $q$; the distribution is defined in an infinite-dimensional space in principle.
In this practical case, which represents the spectrum by a finite number ($N$) of rigidity bins, the Gaussian process is expressed by a multivariate normal distribution for the parameter $\boldsymbol{\theta}$ in the $N$-dimension space, as
\begin{equation}
    \mathcal{P}\qty(\boldsymbol{\theta})
    = \mathcal{N} \qty(\boldsymbol{\theta} \;|\; \boldsymbol{\Theta}_G, \boldsymbol{\Sigma}_G)
    \label{eq:gauss}
\end{equation}
where $\boldsymbol{\Theta}_G$ and $\boldsymbol{\Sigma}_G$ are the mean vector and covariance matrix of the Gaussian process, respectively.
We adopt the radial basis function kernel for the covariance matrix as
\begin{equation}
    \Sigma_{G,kk'} = \sigma_G^2 \exp \qty( -\frac{\qty|q_k - q_{k'}|^2}{b^2} ).
    \label{eq:cov_gauss}
\end{equation}
The diagonal component of $\boldsymbol{\Sigma}_G$ is provided as $\Sigma_{G,kk} = \sigma_G^2$, representing a variance of the parameter $\theta_k$.
The non-diagonal component divided by the variance $\sigma_G^2$, i.e., $\exp \qty(-\qty|q_k - q_{k'}|^2 / b^2)$, represents a correlation coefficient between $\theta_k$ and $\theta_{k'}$, confining an acceptable range of the smoothness of the spectrum.
Detailed descriptions of the conditional probability $\mathcal{P}\qty(\boldsymbol{\eta} | \boldsymbol{\theta})$ and Gaussian process $\mathcal{P}\qty(\boldsymbol{\theta})$ are provided in Appendix \ref{sec:likelihood-gauss} and \ref{sec:hyper}.

Implementing the Gaussian process $\mathcal{P}\qty(\boldsymbol{\theta})$ as a prior probability distribution in Bayesian estimation, the posterior distribution of the parameter $\boldsymbol{\theta}$ is derived as
\begin{equation}
    \mathcal{P} \qty( \boldsymbol{\theta} | \boldsymbol{\eta} ) \propto
    \mathcal{P} \qty( \boldsymbol{\eta} | \boldsymbol{\theta} ) \mathcal{P}\qty(\boldsymbol{\theta}).
    \label{eq:bayes}
\end{equation}
The Gaussian process $\mathcal{P}\qty(\boldsymbol{\theta})$ forms a conjugate prior distribution of the normal distribution $\mathcal{P} \qty( \boldsymbol{\eta} | \boldsymbol{\theta} )$ in this Bayesian estimation.
Therefore, the posterior distribution is also a multivariate normal distribution written as
\begin{equation}
    \mathcal{P} \qty( \boldsymbol{\theta} | \boldsymbol{\eta} ) =
    \mathcal{N} \qty( \boldsymbol{\theta} \;|\; \boldsymbol{\Theta}, \boldsymbol{\Sigma}).
    \label{eq:posterior}
\end{equation}
This equation can be analytically solved, providing another advantage of our method compared to the conventional approach, which generally requires a numerical fitting of parameters, such as the index $\gamma$ for the power-law spectrum.
From Appendix \ref{sec:likelihood-gauss}, the mean vector and covariance matrix in equation (\ref{eq:posterior}) are derived as
\begin{equation}
    \begin{split}
        \boldsymbol{\Theta} &=
        \boldsymbol{\Sigma}_G \qty( \boldsymbol{\Sigma}_L + \boldsymbol{\Sigma}_G )^{-1} \boldsymbol{\Theta}_L
        + \boldsymbol{\Sigma}_L \qty( \boldsymbol{\Sigma}_L + \boldsymbol{\Sigma}_G )^{-1} \boldsymbol{\Theta}_G \qq{and}\\
        \boldsymbol{\Sigma} &=
        \boldsymbol{\Sigma}_L \qty( \boldsymbol{\Sigma}_L + \boldsymbol{\Sigma}_G )^{-1} \boldsymbol{\Sigma}_G.
    \end{split}
    \label{eq:posterior-param}
\end{equation}
The vector $\boldsymbol{\Theta}_L$ and matrix $\boldsymbol{\Sigma}_L$ are parameters of the conditional probability $\mathcal{P}\qty(\boldsymbol{\eta} | \boldsymbol{\theta})$ and derived as
\begin{equation}
    \boldsymbol{\Theta}_L = \boldsymbol{\Sigma}_L \boldsymbol{p} \qq{and}
    \boldsymbol{\Sigma}_L = \boldsymbol{Q}^{-1} \qq{where}
    \boldsymbol{p} = \boldsymbol{C}^\top \boldsymbol{W} \boldsymbol{\eta} \qq{and}
    \boldsymbol{Q} = \boldsymbol{C}^\top \boldsymbol{W} \boldsymbol{C}.
    \label{eq:likelihood-param}
\end{equation}
The weight matrix $\boldsymbol{W}$ is a diagonal matrix with its diagonal elements $W_{ll} = 1/\qty(\sigma_l^{TA})^2$.
In the hyperparameters of the Gaussian process, the mean vector $\boldsymbol{\Theta}_G$ is determined every year, while parameters for the covariance matrix $\boldsymbol{\Sigma}_G$ are set as common values $\sigma_G = 0.1\%$ and $b = 0.4$ [$\log_{10}$(GV)] for all years, as described in Appendix \ref{sec:hyper}.
The marginal distribution of $\mathcal{P}\qty(\boldsymbol{\theta} | \boldsymbol{\eta})$ in each rigidity bin $k$ is derived as
\begin{equation}
    \mathcal{P}\qty(\theta_k | \boldsymbol{\eta})
    = \int \mathcal{P}\qty(\boldsymbol{\theta} | \boldsymbol{\eta}) \dd{\boldsymbol{\theta}_{(k)}'}
    = \int \mathcal{N}\qty(\boldsymbol{\theta} \;|\; \boldsymbol{\Theta}, \boldsymbol{\Sigma}) \dd{\boldsymbol{\theta}_{(k)}'}
    = \mathcal{N}\qty(\theta_k \;|\; \Theta_k, \Sigma_{kk})
\end{equation}
where $\boldsymbol{\theta}_{(k)}'$ is a subset of the vector $\boldsymbol{\theta}$ from which an element $\theta_k$ is removed.
Therefore the 1-$\sigma$ error $\sigma_k$ attached to the mean spectrum $\Theta_k$ is derived from $\boldsymbol{\Sigma}$ as
\begin{equation}
    \sigma_k^2 = \Sigma_{kk}.
    \label{eq:sigma}
\end{equation}
The next section presents the results of the mean spectrum $\boldsymbol{\Theta}$ and its error $\sigma_k$ derived for each year. 

\section{Results}\label{sec:result}
\begin{figure}
    \centering
    \includegraphics[width=0.6\textwidth]{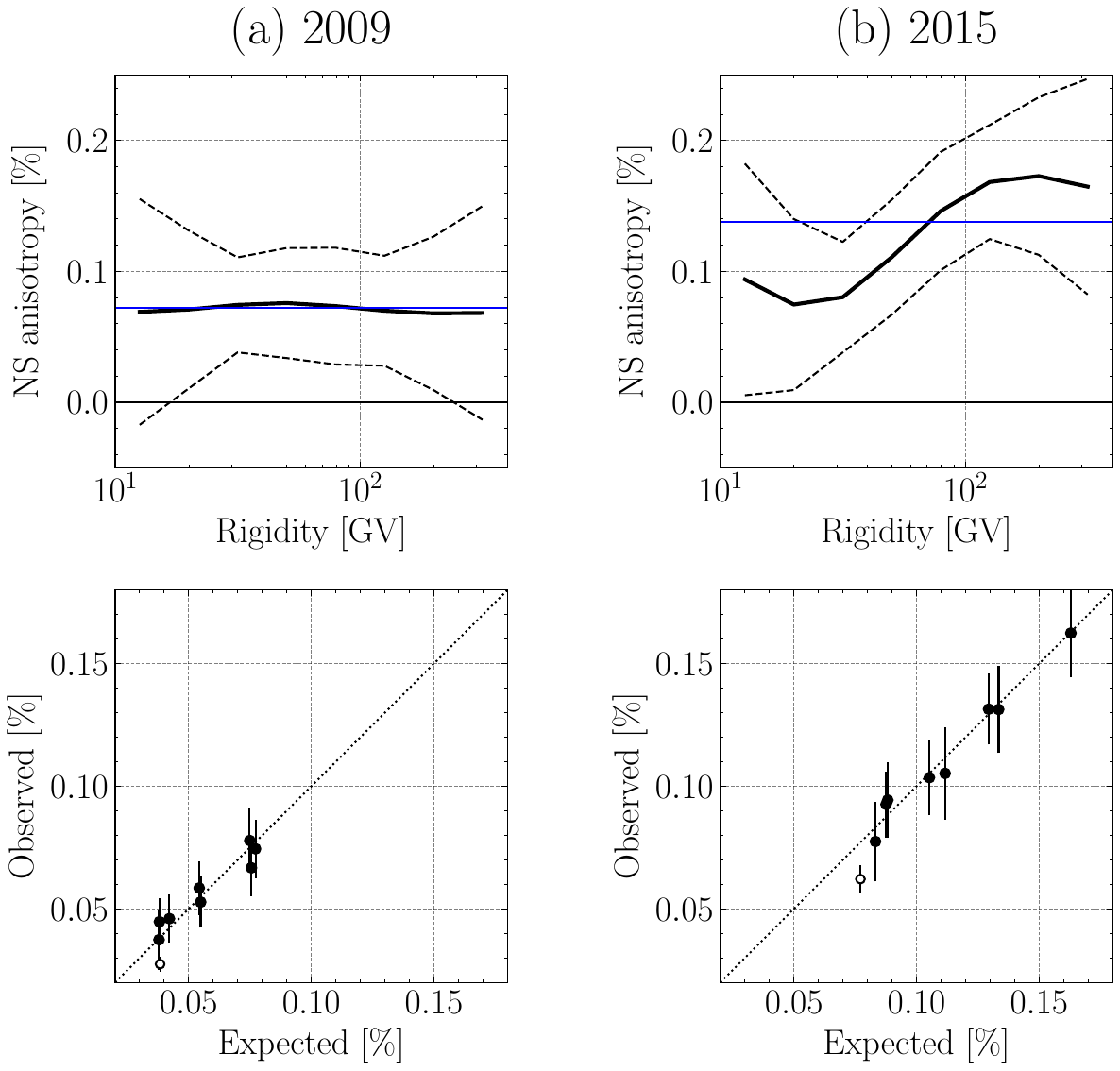}
    \caption{
    Results of Bayesian estimation for the rigidity spectrum of the NS anisotropy [\%] in sample years (a) 2009 and (b) 2015.
    A solid black line in each upper panel is the mean spectrum $\boldsymbol{\Theta}$, while dashed lines above and below the spectrum represent its error range.
    A horizontal blue line is the mean vector $\boldsymbol{\Theta}_G$ of the prior probability distribution.
    Solid circles in each lower panel show a scatter plot between the observable $\eta_l^{TA}$'s of all channel pairs (vertical axis) and their expected values $\tilde\eta_l^{TA}$'s from the mean spectrum (horizontal axis).
    An open circle in each lower panel represents the observed and expected values of Nagoya-GG.
}
    \label{fig:sample_years}
\end{figure}
Figure \ref{fig:sample_years} represents derived rigidity spectra of the NS anisotropy in (a) 2009 on left panels and (b) 2015 on right panels, as sample years around solar activity minimum and maximum, respectively.
A solid black line in each upper panel is the mean spectrum $\boldsymbol{\Theta}$ derived by equation (\ref{eq:posterior-param}), and dashed lines above and below the mean spectrum show its $\pm1\sigma$ error range derived by equation (\ref{eq:sigma}) from observed values $\eta_l^{TA}$'s in each year.
A horizontal blue line is the mean vector $\boldsymbol{\Theta}_G$ of the prior probability distribution.
The mean spectrum represented by the solid black curve seems biased to the prior distribution in the outer rigidity bins ($\sim10$ GV and $\sim300$ GV), resulting in some kinks of the mean spectrum near these rigidities.
This is due to the less responses in these rigidities as shown in Figure \ref{fig:dcpl}, while it negligibly affects scientific discussions, covered by large errors in these rigidity bins.

Solid circles in each lower panel in Figure \ref{fig:sample_years} show a scatter plot between the observable $\eta_l^{TA}$'s of all channel pairs (vertical axis) and their expected values $\tilde\eta_l^{TA}$'s from the mean spectrum (horizontal axis) in each year.
The expected value $\tilde\eta_l^{TA}$ is calculated by replacing $\boldsymbol{\theta}$ in equation (\ref{eq:exp}) with the mean spectrum $\boldsymbol{\Theta}$.
They are consistent with observed values within error bars in each year, ensuring that our new method successfully reconstructs the observed rigidity spectrum of the NS anisotropy with muon counting rates.
An open circle in each lower panel represents the observed and expected values of Nagoya-GG.
The Nagoya-GG is not used to reconstruct the NS anisotropy spectrum in this paper, but its expected value from the derived spectrum shows agreement with its observed value in a comparable range with solid circles.
This result demonstrates that our new analysis method is consistent with the conventional method while providing advanced insights, such as the yearly variation of the anisotropy spectrum.

It is also found that the standard error of Nagoya-GG is smaller than solid circles by a factor less than $\sim1/2$.
One of the causes is that Nagoya-GG double-counts channel pairs in equation (\ref{eq:nagoya-gg}), reducing the statistical error by a factor of $\sim1/\sqrt{2}$.
The remaining factor is likely a local effect different in different stations and remained uncorrected \citep{munakata2023}, representing the limit of our method of directly combining multiple stations.
On the other hand, the solid circles show larger magnitudes than the open circle in each panel, enlarging their significance, thanks to a higher response of each channel pair than Nagoya-GG as shown in Figure \ref{fig:dcpl}.

\begin{figure}
    \centering
    \includegraphics[width=1\textwidth]{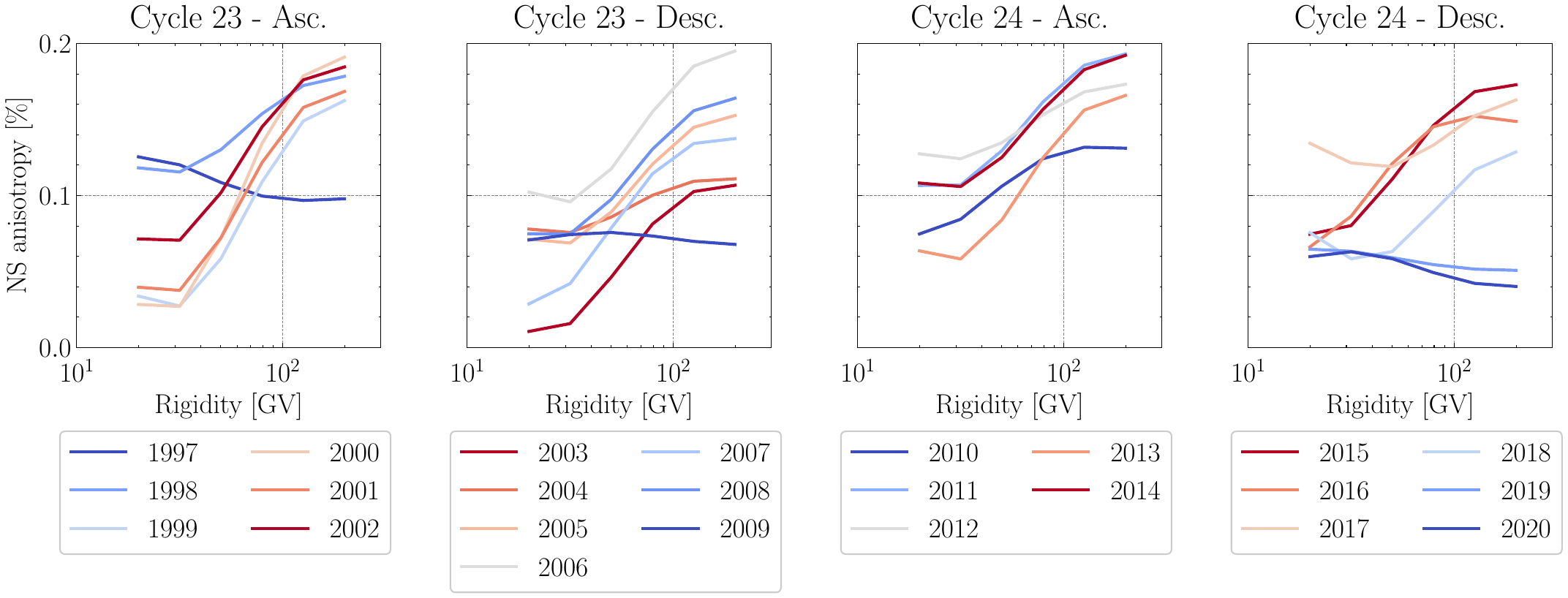}
    \caption{Each panel shows yearly-mean rigidity spectra of the NS anisotropy [\%] in the solar activity's ascending (Asc.) or descending (Desc.) phases in solar cycles 23 or 24, respectively.}
    \label{fig:spectrum}
\end{figure}
The harder slope of the mean spectrum in the solar maximum (2015) than the minimum (2009) in Figure \ref{fig:sample_years} represents a common feature of the solar cycle variation of the NS anisotropy, as demonstrated by Figure \ref{fig:spectrum}.
It displays the mean spectrum every year from 1997 to 2020, split into the solar-activity ascending and descending phases in the solar cycle 23 or 24 in each panel.
The edge rigidity bins around $\sim10$ GV and $\sim300$ GV are truncated in this figure, considering their relative unreliability mentioned above.
Each year is denoted by a legend below each panel, and the gradation of line colors indicates the solar activity, where the red and blue colors correspond to years around the activity maximum and minimum, respectively.
Overall, the blue lines have a softer slope than the red lines in each panel, indicating that the NS anisotropy exhibits a softening of its rigidity spectrum in the solar activity minimum.

\cite{yasue1980} also suggested a variation of the NS anisotropy's rigidity spectrum according to the solar activity, and since then, quantitative estimation of the spectrum has been rarely reported.
\cite{yasue1980} provided only two average spectra around the maximum and declining solar activity phases, each in 1969-1970 and 1971-1973, and concluded an insignificant difference of the spectrum between the two periods.
In that analysis, multiple types of observations, including underground muon detectors, were required, and the power-law rigidity spectrum of the anisotropy was presumed.
These limitations probably prevented the study from revealing the dynamic variation of the anisotropy with a better temporal resolution.
On the other hand, our approach successfully reconstructs the spectrum every year, as shown in Figure \ref{fig:spectrum}, thanks to the better tolerance of varying spectral shapes.
A physical origin of the revealed variation of the spectrum is discussed in Section \ref{subsec:modparam}.

\section{Discussions}
\subsection{Modulation parameters}\label{subsec:modparam}
The sector reversal of the NS anisotropy is expected as a consequence of the diamagnetic drift of GCRs with the radial density gradient $g(P)$ in the IMF, expressed as \citep{yasue1980}
\begin{equation}
    \frac{\xi_z^{TA}(P)}{\cos\delta_E} \sim R_L\, g(P) \sin\psi = \frac{P}{cB} g(P) \sin\psi.
    \label{eq:xi-g}
\end{equation}
The radial density gradient $g(P)$ is related to the GCR density as
\begin{equation}
    g(P) = \frac{1}{U(r_E,P)}
    \left. \frac{\partial U(r,P)}{\partial r} \right|_{r=r_E}
\end{equation}
where $U(r)$ is a GCR density at the distance $r$ from Sun and $r_E$ is the position of Earth.
In equation (\ref{eq:xi-g}), it is approximated that the density gradient of GCRs in the ecliptic plane is dominated by its radial component, especially in the Bartels rotation mean, which averages the azimuthal (GSE-$y$) component out to zero.
The gyro-radius $R_L$ of GCRs is derived as $R_L = P/(cB)$ from the GCR rigidity $P$, IMF magnitude $B$ and light speed $c$.
Derivation of the NS anisotropy in Section \ref{sec:method} is based on the GEO coordinate system, and $\xi_z^{TA}$ is expected to be a projection of the sector reversal of the GSE-$z$ component of the anisotropy onto the GEO-$z$ axis.
Therefore, the NS anisotropy is divided by the factor $\cos\delta_E$, where $\delta_E$ is the inclination angle of Earth's rotation axis from GSE-$z$ axis, in equation (\ref{eq:xi-g}) to convert the NS anisotropy $\xi_z^{TA}$ from the GEO to GSE coordinate system.

Referring to the derivation procedure of $\eta_{ij}^{TA}$ in Section \ref{subsec:eta}, GSE-$x$ and $y$ components of the IMF vector are calculated from OMNIWeb data as
\begin{equation}
    B_{x(y)}^{TA} = \frac{B_{x(y)}^T - B_{x(y)}^A}{2}
\end{equation}
where $B_{x(y)}^T$ and $B_{x(y)}^A$ are averages of the GSE-$x(y)$ component of the IMF in the $Toward$ and $Away$ sectors respectively in each year.
The IMF magnitude and spiral angle $\psi$ in equation (\ref{eq:xi-g}) are consequently derived as
\begin{equation}
    B = \sqrt{\qty(B_x^{TA})^2 + \qty(B_y^{TA})^2} \qq{and} \sin \psi = - \frac{B_y^{TA}}{B}.
    \label{eq:b}
\end{equation}

Replacing $\xi_z^{TA}(P)$ in equation (\ref{eq:xi-g}) with the mean spectrum $\Theta_k$ derived in Section \ref{subsec:bayes}, the most probable value of the radial density gradient in each rigidity bin $k$ is derived as
\begin{equation}
    g_k \sim - \frac{c}{P_k^m \cos\delta_E} \cdot \frac{B^2}{B_y^{TA}} \Theta_k
    \label{eq:g-xi}
\end{equation}
where the rigidity $P$ is represented by its median $P_k^m$ in each rigidity bin.
A square of its error is estimated as
\begin{equation}
\begin{split}
    \sigma^2(g_k) &= \qty{ \frac{\partial g_k}{\partial B_x^{TA}} \sigma\qty(B_x^{TA}) }^2 +
               \qty{ \frac{\partial g_k}{\partial B_y^{TA}} \sigma\qty(B_y^{TA}) }^2 +
               \qty( \frac{\partial g_k}{\partial \Theta_k} \sigma_k )^2\\
    &= \qty( \frac{c}{P_k^m \cos\delta_E B_y^{TA}} )^2
    \qty[ \qty{ 2 B_x^{TA} \Theta_k \sigma\qty(B_x^{TA}) }^2
    + \qty{ \qty( B_y^{TA} - \frac{\qty(B_x^{TA})^2}{B_y^{TA}} ) \Theta_k \sigma\qty(B_y^{TA}) }^2
    + \qty( B^2 \sigma_k )^2 ]
\end{split}
\end{equation}
where $\sigma\qty(B_x^{TA})$ and $\sigma\qty(B_y^{TA})$ are errors of $B_x^{TA}$ and $B_y^{TA}$ respectively.
Similarly to the error of $\eta_{ij}^{TA}$ in equation (\ref{eq:sigma-eta}), these errors are derived as
\begin{equation}
    \sigma^2\qty(B_{x(y)}^{TA}) = \frac{\sigma^2\qty(B_{x(y)}^T) + \sigma^2\qty(B_{x(y)}^A)}{4}
\end{equation}
where $\sigma\qty(B_{x(y)}^T)$ and $\sigma\qty(B_{x(y)}^A)$ are derived from standard deviations of the GSE-$x$($y$) component of the daily IMF in the $Toward$ and $Away$ sectors respectively.

\begin{figure}
    \centering
    \includegraphics[width=1\textwidth]{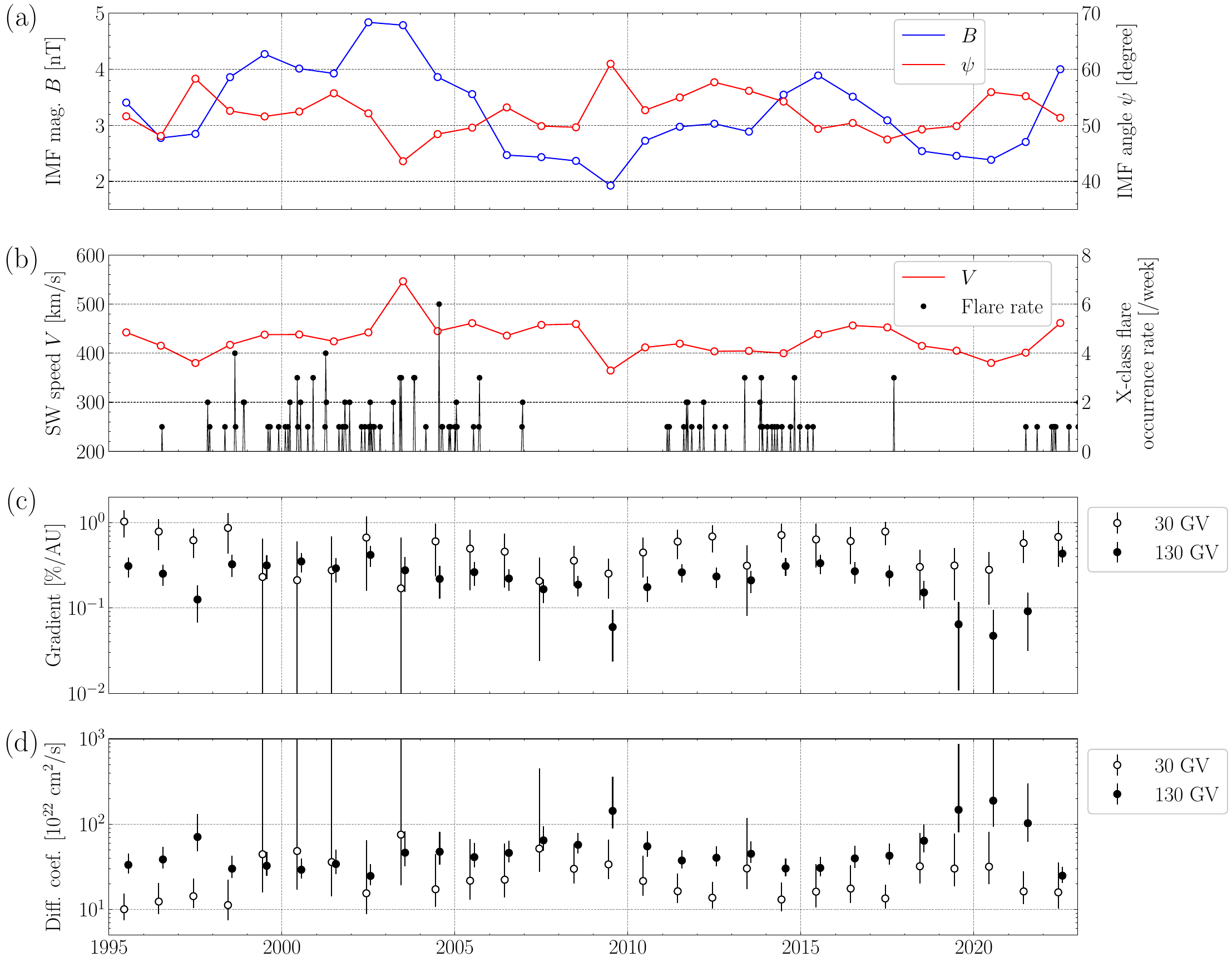}
    \caption{
    Temporal variations of the solar activity, interplanetary, and GCR modulation parameters.
    (a) Yearly averages of the magnitude $B$ [nT] and spiral angle $\psi$ of the IMF.
    (b) Solar wind (SW) speed $V$ [km/s] for each year and X-class flare occurrence rate every week.
    (c) Radial density gradients [\%/AU] and (d) diffusion coefficients [10$^{22}$ cm$^2$/s] of (open circles) $\sim30$ GV and (solid circles) $\sim130$ GV GCRs derived from the yearly NS anisotropy.}
    \label{fig:modparam}
\end{figure}
To reveal a solar cycle variation of the radial density gradient along with its rigidity dependence, we extract $g_k$'s in $k=3$ and $k=6$ bins corresponding to the rigidity of $P\sim30$ GV and $P\sim130$ GV, respectively.
These rigidity bins are selected not to be affected by significant errors in rigidity bins below $\sim20$ GV and above $\sim200$ GV, shown by upper panels of Figure \ref{fig:sample_years}.
Figure \ref{fig:modparam}c displays a yearly variation of the density gradient in each rigidity bin, in a unit of \%/AU and a logarithmic vertical axis.
The solar activity and interplanetary states are represented by Figures \ref{fig:modparam}a and \ref{fig:modparam}b;
the panel (a) displays yearly averages of the IMF magnitude $B$ (blue line) and spiral angle $\psi$ (red line) derived in equation (\ref{eq:b}), red line in the panel (b) shows yearly averages of the solar wind speed $V$, and black points in the panel (b) are occurrence rate of the X-class solar flares every week.
In the same manner as the IMF, the yearly average of the solar wind speed is derived as $V = (V^T + V^A)/2$ where $V^T$ and $V^A$ are averages of the solar wind speed in the OMNIWeb for the $Toward$ and $Away$ sectors, respectively, every year.
The IMF spiral angle $\psi$ and solar wind speed $V$ show an anti-correlation with each other, following the Parker spiral picture.
The X-class flare occurrence rate is derived from the Konus-WIND flare catalog \citep{konus-windteam2024,palshin2014}.

We can identify a solar cycle variation of the radial density gradient at $P\sim130$ GV (solid circles in Figure \ref{fig:modparam}c), where the density gradient is suppressed in the solar activity minima around 1997, 2009, and 2020.
The density gradient at $P\sim30$ GV (open circles in Figure \ref{fig:modparam}c) also shows minima at least around 2009 and 2020, but its relative variation expressed in the logarithmic scale is smaller than that of $P\sim130$ GV.
We can conclude that the solar cycle variation of the NS anisotropy's rigidity spectrum as seen in Figure \ref{fig:spectrum} is attributed to this rigidity-dependent solar cycle variation of the density gradient.

Despite the hard spectrum of the NS anisotropy overall in Figure \ref{fig:spectrum}, the radial density gradient at $P\sim30$ GV is generally larger than $P\sim130$ GV in Figure \ref{fig:modparam}c, indicating a soft spectrum of the density gradient.
This soft spectrum is reasonable from a picture of the smaller solar modulation in higher rigidity GCRs.
From a technical viewpoint, it is caused by the factor $1/P_k^m$ multiplied in equation (\ref{eq:g-xi}).
While a quantitative reproduction of the observed anisotropy has been rarely performed by heliospheric simulations except for a small number of studies \citep[e.g.,][]{kadokura1986b}, the radial density gradient, or radial distribution of GCR density, is predicted by some simulation works for a realistic heliosphere model including the termination shock and heliosheath \citep{langner2003,langner2004,ngobeni2010}. 
Our result on the radial density gradient will provide an observational validation of such heliospheric models.

It is also worth focusing on some years deviating from the general picture of the density gradient with a soft spectrum, as seen in 1999-2001, 2003, 2007, and 2013 in Figure \ref{fig:modparam}c.
While being insignificant because of the large errors, these years have rapid suppression of the density gradient only in $\sim30$ GV GCRs from the previous years, resulting in comparable density gradients between $\sim30$ GV and $\sim130$ GV GCRs.
Such a hard spectrum of the density gradient causes a noticeably hard spectrum of the NS anisotropy in these years, as displayed in Figure \ref{fig:spectrum}.
Most of these years, 1999-2001, 2003, and 2013, correspond to the years when the area or number of low-latitude coronal holes was enhanced on Sun, as seen in the results of automatic identification of the coronal holes \citep{andreeva2022,fujiki2016}.
The remaining year, 2007, is just after the coronal mass ejection (CME) event on December 13th, 2006, which was the biggest halo CME since the ``Halloween storm'' in 2003 as of the study by \cite{liu2008}.
From black points in Figure \ref{fig:modparam}b, it is found that a few X-class flare events, including the CME on December 13th, 2006, successively occurred after a relatively calm period without X-class flares in 2005-2006. 
A similar situation was found in the last half of 2017, where three successive X-class flares were detected after over $\sim2$ years with no X-class flares.
After this event, the density gradient shows a spectral hardening in 2017-2018, similar to 2006-2007, as displayed by the open and solid circles getting closer to each other in these years in Figure \ref{fig:modparam}c.
These results imply that the post-storm interplanetary state after discrete solar eruptions or the low-latitude coronal holes mentioned above possibly caused unusual modulation effects suppressing the density gradient in lower rigidity GCRs, although more detailed analyses are required.

Assuming a radial balance between the diffusion flow and solar wind convection of GCRs further provides an estimation of their diffusion coefficient, as
\begin{equation}
    \kappa_{rr}(P) g(P) \sim \frac{2 + \Gamma}{3} V
    \label{eq:diff-conv}
\end{equation}
where $\Gamma \sim 2.7$ is a power-law index of the energy spectrum of the GCR density.
The radial diffusion coefficient $\kappa_{rr}$ of GCRs at the position of Earth is expressed as
\begin{equation}
    \kappa_{rr} = \kappa_\parallel \cos^2 \psi + \kappa_\perp \sin^2 \psi
    \label{eq:kpara-kperp}
\end{equation}
where $\kappa_\parallel$ and $\kappa_\perp$ are diffusion coefficients parallel and perpendicular to the IMF.
Therefore we can estimate the radial diffusion coefficient $\kappa_{rr,k}$ in each rigidity bin $k$ as
\begin{equation}
    \kappa_{rr,k} \sim \frac{2 + \Gamma}{3} \cdot \frac{V}{g_k}.
\end{equation}
The error range of $\kappa_{rr,k}$ can no longer be approximated by a symmetric normal distribution because it is dominated by the error of $g_k$, which is inversely proportional to $\kappa_{rr,k}$.
We approximate the lower and upper limits of a $1\sigma$ confidence interval of $\kappa_{rr,k}$ by
\begin{equation}
    \kappa_{rr,k}^- = \frac{2 + \Gamma}{3} \cdot \frac{V}{g_k + \sigma(g_k)} \qq{and}
    \kappa_{rr,k}^+ = \frac{2 + \Gamma}{3} \cdot \frac{V}{g_k - \sigma(g_k)}
    \label{eq:sigma_krr}
\end{equation}
respectively.
A standard deviation of the solar wind speed $V$ is ignored because it is negligible compared to $\sigma\qty(g_k)$.
Figure \ref{fig:modparam}d displays temporal variations of the radial diffusion coefficient $\kappa_{rr}$ for $\sim30$ GV and $\sim130$ GV GCRs.
It demonstrates a solar cycle variation of the diffusion coefficient in both rigidities as well as the radial density gradient but negatively correlates with solar activity.
We have to note that the diffusion and convection balance in equation (\ref{eq:diff-conv}), known as the force-field approximation \citep{gleeson1968,gleeson1973}, is based on the diurnal anisotropy's phase observed to be around 18:00 local solar time \citep{rao1963}.
However, \cite{chen1993} and \cite{munakata2014} showed a 22-year cycle excursion of the phase to earlier local time, suggesting that the left side of equation (\ref{eq:diff-conv}) can be $\sim4/5$ of the right side in the case of such an excursion.
Therefore, this estimation of the diffusion coefficient $\kappa_{rr}$ can have a systematic error of $\sim20\%$ in addition to equation (\ref{eq:sigma_krr}).

\begin{figure}
    \centering
    \includegraphics[width=0.6\textwidth]{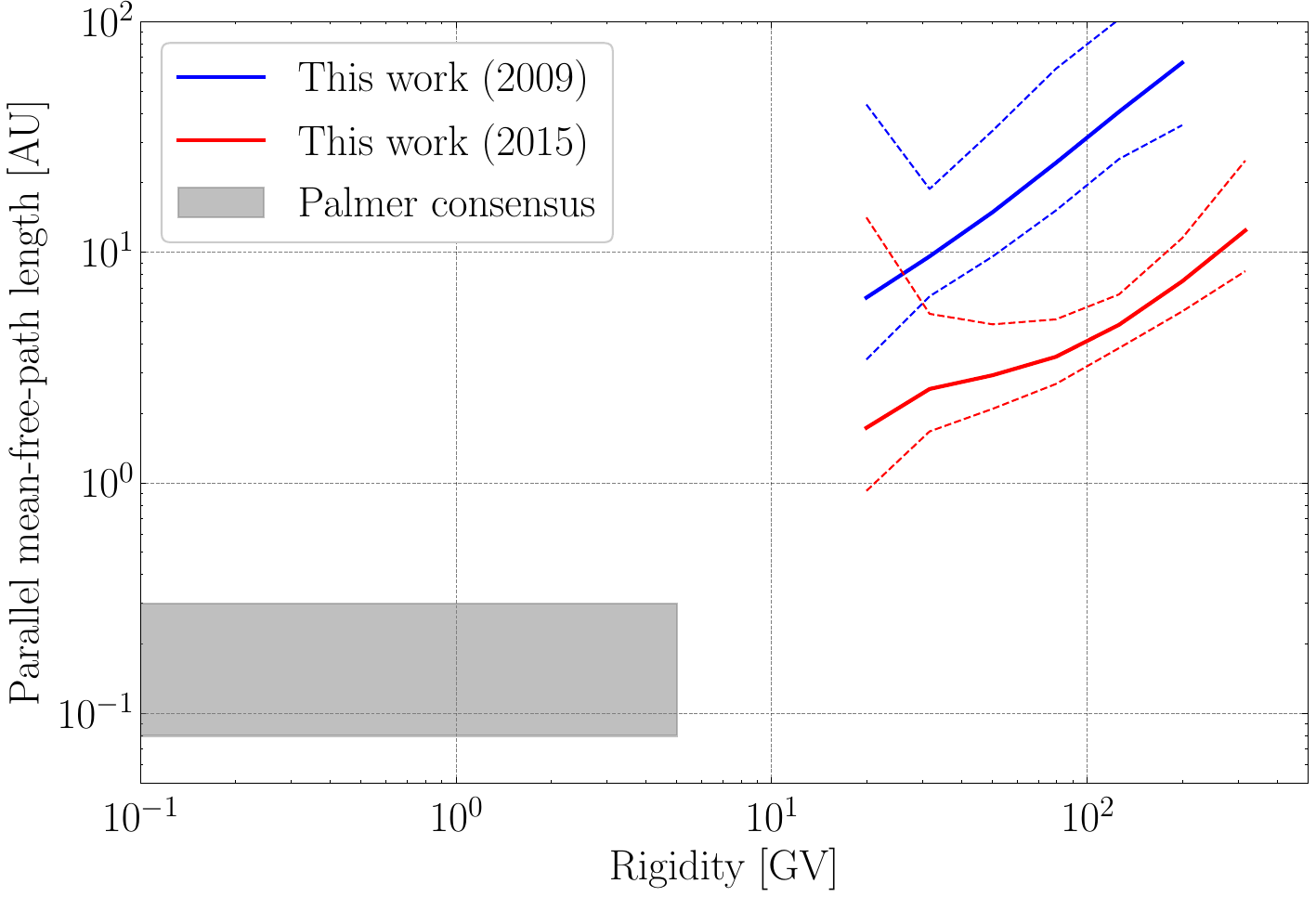}
    \caption{Rigidity spectrum of the parallel mean-free-path length [AU] of GCRs for 2009 (solar activity minimum) and 2015 (maximum).
    The gray shaded area is ``Palmer consensus'' region mainly by SEP observations.}
    \label{fig:mfp}
\end{figure}
Based on a comparable $\sin^2\psi$ with $\cos^2\psi$ in the average IMF and $\kappa_\perp/\kappa_\parallel \ll 1$ supported by previous studies \citep[e.g.,][]{burger2000,engelbrecht2022}, we obtain the parallel diffusion coefficient as
\begin{equation}
    \kappa_{\parallel,k} \sim \frac{\kappa_{rr,k}}{\cos^2\psi}
\end{equation}
from equation (\ref{eq:kpara-kperp}) for the rigidity bin $k$.
Therefore, the parallel mean-free-path length of GCRs and its error range are estimated as
\begin{equation}
    \lambda_{\parallel,k} = \frac{3\kappa_{\parallel,k}}{v} \sim \frac{3\kappa_{rr,k}}{v\cos^2\psi},\quad
    \lambda_{\parallel,k}^- \sim \frac{3\kappa_{rr,k}^-}{v\cos^2\psi},\qq{and}
    \lambda_{\parallel,k}^+ \sim \frac{3\kappa_{rr,k}^+}{v\cos^2\psi}
\end{equation}
where $v \sim c$ is the particle velocity and approximated by a light speed $c$ for relativistic GCRs with $P > 10$ GV.
Blue and red curves in Figure \ref{fig:mfp} display rigidity spectra of the parallel mean-free-path length in 2009 and 2015, respectively, as sample years in the solar activity minimum and maximum.
The outer rigidity bins with substantial errors are truncated in this figure.
The gray shaded area is ``Palmer consensus'' region which was proposed by \cite{palmer1982} and has been validated mainly by solar energetic particle (SEP) observations and numerical simulations \citep[e.g.,][]{bieber1994,tautz2013,engelbrecht2022}.
Our result extends the mean-free-path length estimation into higher rigidity region $\le200$ GV, and from Figure \ref{fig:mfp} it is expected to smoothly connect to the consensus region by extrapolating the rigidity dependence.

Analyses of the GCR density and its rigidity spectrum also provide estimations of the diffusion coefficient or mean-free-path length in $>10$ GV region \citep{kojima2024,tomassetti2023} by surveying optimal heliospheric parameters fitting to observations.
Our result of the mean-free-path length in Figure \ref{fig:mfp} is consistent with their results in the order of magnitude, while our result features the unprecedented temporal resolution on a yearly basis and the wide rigidity range of over one order of magnitude.
It is also noted that their approach using the GCR density requires the assumption on the whole heliospheric structure, such as a spatial distribution of the diffusion coefficient, because the GCR density at Earth reflects an integration of the modulation effects from the outer edge of the heliosphere.
On the other hand, the NS anisotropy can provide an estimation of the diffusion coefficient free from such an assumption as well as the SEP observations in $<10$ GV region.
This is one of the advantages of the anisotropy observation, leading to a relatively reliable estimation of the diffusion coefficient.

\subsection{Future prospects of the NS conjugate observation}\label{subsec:polar}
\begin{figure}
    \centering
    \includegraphics[width=1\textwidth]{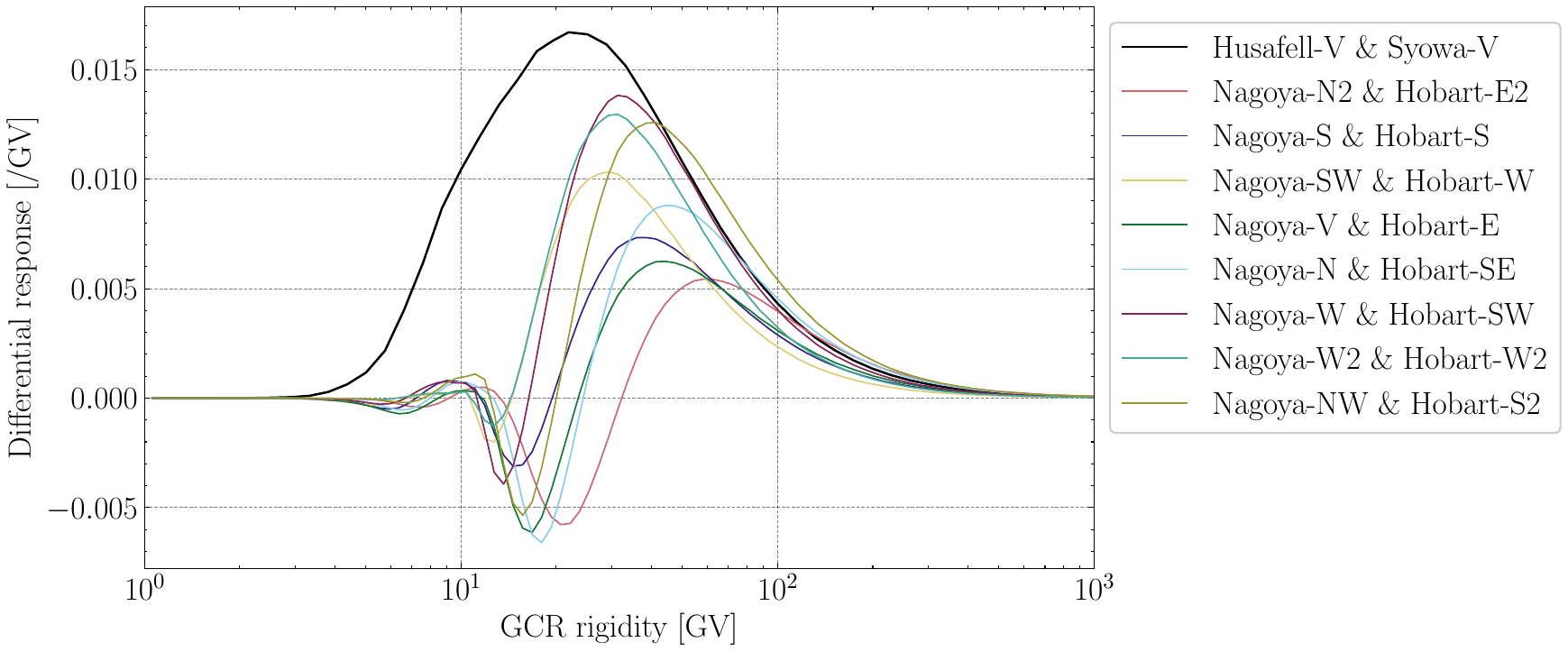}
    \caption{Black bold line is a differential response to the NS anisotropy for vertical channels of muon detectors at Husafell, Iceland (virtual detector) and Syowa Station. Other solid lines are repeats of those in Figure \ref{fig:dcpl}.}
    \label{fig:ice-syowa}
\end{figure}
Recently, we started a muon observation at Syowa Station, Antarctica, in 2018 in collaboration with the National Institute of Polar Research (NIPR), Japan \citep{kato2021a}.
NIPR also has a collaborative research framework with Iceland, located around an Arctic geomagnetic conjugate point with Syowa Station.
The polar conjugate observation in Iceland and Syowa Station will be an expanded concept of the current Nagoya-Hobart observation, and we provide a result of its quantitative simulation in this section.
The black bold line in Figure \ref{fig:ice-syowa} displays the differential response to the NS anisotropy for vertical channels of muon detectors at Husafell, Iceland, and Syowa Station.
The virtual detector at Husafell is set to have an equivalent geometry with the Syowa detector having a 1$\times$2 m$^2$ detection area.
Other lines in Figure \ref{fig:ice-syowa} are the responses of channel pairs used in this study, i.e., repeats of those in Figure \ref{fig:dcpl}.
Notably, only the Husafell-Syowa pair has a substantial response in the lower rigidity ($<20$ GV) region, without a polarity reversal into a negative response seen in other channel pairs.
The north- and southward inclined channels in Nagoya and Hobart detectors respectively view high-latitude directions, but the geomagnetic field around the mid-latitude stations deflects cosmic-ray trajectories into lower-latitude directions as briefly discussed in Section \ref{subsec:bayes}.
This effect is more substantial for lower-rigidity cosmic rays and prevents us from gaining a significant response to the NS anisotropy in lower-rigidity regions only by mid-latitude stations.
Arctic and Antarctic regions, such as Iceland and Syowa Station, are relatively free from this effect and are predicted to allow us to observe the NS anisotropy in lower-rigidity GCRs accurately.
While having less angular resolutions than muon detectors, neutron monitors in Arctic and Antarctic regions are also expected to be responsive to the NS anisotropy in further lower rigidity ($\sim1$ GV) regions.
Therefore, the concept of Iceland-Syowa conjugate muon observation can be a bridge to connect the muon and neutron detector networks.

\section{Conclusion}
A new analysis method to derive the rigidity spectrum of GCR anisotropy from ground-based observations has been developed and demonstrated by revealing a yearly variation of the NS anisotropy's spectrum.
In this method, atmospheric temperature effects on cosmic-ray muons are directly corrected by the meteorological reanalysis data, allowing for combining multiple muon detectors in a network observation free from individual local effects.
General graph matching in graph theory is adopted to survey optimal combinations of directional channels, which ensures high sensitivity to the NS anisotropy.
The highlight of our analysis method is Bayesian estimation with the Gaussian process, which has the potential to be applied to any unfolding problem of the ground-based observations to derive GCR properties in space.
The Gaussian process only confines the acceptable ranges of the spectrum value and its smoothness without supposing any analytical function, providing a sufficient tolerance of varying spectral shapes to trace dynamic variations of the GCR spectrum in solar modulation phenomena.

Previous works deriving rigidity spectra of the anisotropy use simultaneous observation data by multiple detector types, i.e., ground-based muon detectors, underground muon detectors, and neutron monitors \citep{yasue1980,hall1994,munakata1997,munakata2022}.
On the other hand, this paper succeeded in deriving the NS anisotropy's spectrum only by ground-based muon detectors in Nagoya and Hobart.
This relaxed requirement for observational data will make the anisotropy analysis a more common approach in cosmic-ray research.
The minimized number of observations also ensures a uniform dataset for an extended period, allowing this study to reveal the solar cycle variation of the NS anisotropy's spectrum.
Softening of the spectrum in the solar activity minima was discovered, and it is attributed to the rigidity-dependent variation of the radial density gradient of GCRs.
The diffusion coefficient or mean-free-path length of GCRs is subsequently derived based on the force-field approximation, and it is demonstrated that our analysis expands the mean-free-path length estimation into $\le200$ GV region from $<10$ GV region achieved by SEP observations.
The rigidity-dependent diffusion coefficient has been a critical problem in elucidating the GCR propagation \citep{engelbrecht2022}, emphasizing the importance of the anisotropy observation and our result.

In addition to expanding the muon detector network into polar regions as described in Section \ref{subsec:polar}, applying our analysis scheme to a broader range of cosmic-ray studies is desired.
Reconstruction of the three-dimensional anisotropy or density spectrum of GCRs is one such scientific target, and it can, in principle, be achieved by generalizing the Gaussian process used in this paper.
Short-term disturbance phenomena, including CME events, are also crucial topics along with the solar cycle variation analyzed in this paper, and an application of our analysis method to such an event will be presented in the future.

\begin{acknowledgements}
This research was supported by the ``Strategic Research Projects'' grant from ROIS (Research Organization of Information and Systems) and by JSPS KAKENHI Grant Number JP22KK0049.
The GMDN project is partially supported by ROIS-DS-JOINT (030RP2023), JARE AJ1007 program, and JSPS KAKENHI Grant Number JP24K07068.
The observations with Nagoya and Hobart muon detectors are supported by Nagoya University and Australian Antarctic Division, respectively.
\end{acknowledgements}

\appendix
\section{Dataset preparation}\label{sec:prepare}
In this appendix, we describe the preparation procedure of muon counting-rate data performed before the analysis procedure in the main text, mainly the correction for atmospheric temperature effect on muons.
The temperature correction method by \cite{mendonca2016,mendonca2019} defines a mass-weighted temperature for each station as
\begin{equation}
    H_{\rm st} = \sum_{h=1}^{h_{max}-1} w_h \frac{T_h + T_{h+1}}{2}
\end{equation}
where $T_h$ is the atmospheric temperature at altitude $h$ above the location of each station and $h_{max}$ indicates the top of the atmosphere.
The weight $w_h$ is derived as $w_h = (z_h - z_{h+1})/z_0$ where $z_h$ is the atmospheric depth at the altitude $h$ and $z_0 = z_1 - z_{h_{max}}$.
Muon counting rates $N_i(t)$ corrected for atmospheric pressure effects on each time $t$ are published by \cite{gmdncollaboration2024}.
The atmospheric temperature effect $\Delta N_i(t)$ on $N_i(t)$ is expressed as 
\begin{equation}
    \Delta N_i(t) = \alpha_i \qty{ H_{\rm st}(t) - H_{\rm st}^0 }
\end{equation}
for the directional channel $i$ belonging to each station.
The constant $H_{\rm st}^0$ is an average level of the mass-weighted temperature and set at 253 or 250 [K] for Nagoya or Hobart station, respectively.
The coefficient $\alpha_i$ for each directional channel is derived by \cite{hayashi2024} and ranges from -0.28 to -0.24 [\%/K] in Nagoya station's channels and from -0.24 to -0.22 [\%/K] in Hobart station's channels. 
The temperature-corrected counting rate is derived as
\begin{equation}
    N_i^{\rm corr}(t) = N_i(t) - \Delta N_i(t).
    \label{eq:corr_temp}
\end{equation}

The altitude profile of the atmospheric temperature is provided as meteorological reanalysis data by GDAS (Global Data Assimilation System) from 2005 and NCEP/NCAR (National Centers for Environmental Prediction and National Center for Atmospheric Research) until 2004, both published in NOAA's Air Resources Laboratory website \citep{noaaarl2024}.
It provides the altitude and temperature $T_h$ at each isobaric surface from 20 hPa to 1000 hPa at a designated location.
We set the ground-level altitude, $h=1$, at the isobaric surface with the lowest altitude above 500 m.
The top of the atmosphere, $h_{max}$, is approximated by the highest isobaric surface in the data, corresponding to the pressure of 20 hPa.
In calculating the weight $w_h$, the atmospheric pressure of each isobaric surface is used in place of the atmospheric depth $z_h$.
Temporal resolutions of the meteorological reanalysis data are 3 hours in GDAS data and 6 hours in NCEP/NCAR data, respectively, which are insufficient for hourly muon counting-rate data.
We first calculate the mass-weighted temperature $H_{\rm st}(t)$ on a 3-hour or 6-hour basis using the provided data.
Then, it is interpolated by a linear interpolation between timestamps, and hourly $H_{\rm st}(t)$ for the temperature correction is derived.

The corrected muon counting rate is converted into a deviation from its Bartels rotation average,
\begin{equation}
	I_i(t) = \frac{N_i^{\rm corr}(t)}{N_i^{\rm BR}}
\end{equation}
where $N_i^{\rm{BR}}$ is $N_i^{\rm corr}(t)$ averaged over each Bartels rotation.
Hourly counting rates $>2\%$ below or above the Bartels rotation average $N_i^{\rm BR}$ are omitted to eliminate disturbance events such as the Forbush decrease.

\section{Sensitivity of each channel pair to the NS anisotropy}\label{sec:sense}
Optimization of the channel pairing described in Section \ref{subsec:pair} is based on the sensitivity $a_{ij}$ defined for each channel pair $ij$.
In this appendix, we describe its derivation procedure.
From the discussion with equation (\ref{eq:eta-xi}) in Section \ref{subsec:eta}, the perturbation on $\eta_{ij}^{TA}$ from anisotropy components other than the NS anisotropy is estimated as
\begin{equation}
    \Delta\eta_{ij}^{TA} = \int_{P=0}^\infty
 \qty( \epsilon_0^{TA} \dd{c_{ij}^0} + \epsilon_c^{TA} \dd{c_{ij}^d} + \epsilon_s^{TA} \dd{s_{ij}^d} ).
\label{eq:perturb}
\end{equation}
Previous studies \citep{lockwood1960,yasue1980,munakata1997,kozai2016} report nearly flat spectra for the anisotropy components and a soft spectrum proportional to $\sim P^{-1}$ for the short-term density variations.
Based on these suggestions, we temporarily adopt an ad hoc simplification on the rigidity spectra only for the channel pairing as
\begin{equation}
    \xi_z^{TA}(P) = \xi_z^{TA},\quad
    \epsilon_0^{TA}(P) = \epsilon_0^{TA}\qty(\frac{P}{60[{\rm GV}]})^{-1},\quad
    \epsilon_c^{TA}(P) = \epsilon_c^{TA},\qq{and}
    \epsilon_s^{TA}(P) = \epsilon_s^{TA}.
\end{equation}
The constant 60 GV for the density variation $\epsilon_0^{TA}(P)$ is introduced as a representative rigidity of muon detectors.
Equations (\ref{eq:eta-xi}) and (\ref{eq:perturb}) are simplified as
\begin{equation}
    \eta_{ij}^{TA} \sim \xi_z^{TA} c_{ij}^z\qq{and}
    \Delta\eta_{ij}^{TA} \sim \epsilon_0^{TA} c_{ij}^0 + \epsilon_c^{TA} c_{ij}^d + \epsilon_s^{TA} s_{ij}^d
 \label{eq:eta-flat}
\end{equation}
where $c_{ij}^z$, $c_{ij}^0$, $c_{ij}^d$, and $s_{ij}^d$ are coupling coefficients for the assumed spectra, defined as
\begin{equation}
    c_{ij}^z = \int_{P=0}^\infty \dd{c_{ij}^z},\quad
    c_{ij}^0 = \int_{P=0}^\infty \qty(\frac{P}{60[{\rm GV}]})^{-1} \dd{c_{ij}^0},\quad
    c_{ij}^d = \int_{P=0}^\infty \dd{c_{ij}^d},\qq{and}
    s_{ij}^d = \int_{P=0}^\infty \dd{s_{ij}^d}.
\end{equation}
From equation (\ref{eq:eta-flat}), the optimized pairs can be defined as those maximizing the response $\partial\eta_{ij}^{TA}/\partial\xi_z^{TA} \sim c_{ij}^z$ to the NS anisotropy while minimizing the response to other anisotropy components which is estimated as
\begin{equation}
    \qty(\frac{\partial\Delta\eta_{ij}^{TA}}{\partial\epsilon_0^{TA}})^2 +
    \qty(\frac{\partial\Delta\eta_{ij}^{TA}}{\partial\epsilon_c^{TA}})^2 +
    \qty(\frac{\partial\Delta\eta_{ij}^{TA}}{\partial\epsilon_s^{TA}})^2
    \sim \qty(c_{ij}^0)^2 + \qty(c_{ij}^d)^2 + \qty(s_{ij}^d)^2.
\end{equation}
Therefore, we define the sensitivity of each channel pair $ij$ to the NS anisotropy as
\begin{equation}
    a_{ij} = \left\{\begin{array}{ll}
        c_{ij}/\sqrt{\qty(c_{ij}^0)^2 + \qty(c_{ij}^d)^2 + \qty(s_{ij}^d)^2} & \qq{for} c_{ij}^z > 0.7 \qq{and} a_{ij} > 1.6\\
        0 & \qq{for} c_{ij}^z \le 0.7 \qq{or} a_{ij} \le 1.6.
    \end{array}\right.
    \label{eq:sense}
\end{equation}
The thresholds, $c_{ij}^z > 0.7$ and $a_{ij} > 1.6$, are introduced to truncate channel pairs with worse sensitivities than Nagoya-GG, whose response and sensitivity are $c_{ij}^z=0.63$ and $a_{ij}=1.51$ respectively.

\section{Likelihood function and Gaussian process for the spectrum parameters}\label{sec:likelihood-gauss}
In equation (\ref{eq:likelihood}), the conditional probability distribution $\mathcal{P}\qty(\boldsymbol{\eta} | \boldsymbol{\theta})$ of observed values $\eta_l^{TA}$'s for the parameter $\boldsymbol{\theta}$ is equivalent to a likelihood function for $\boldsymbol{\theta}$, which is generally used in the maximum likelihood estimation.
The parameters $\boldsymbol{\Theta}_L$ and $\boldsymbol{\Sigma}_L$ in equations (\ref{eq:posterior-param}) and (\ref{eq:likelihood-param}) are respectively identical to the mean vector and covariance matrix of the likelihood function expressed in a multivariate normal distribution.
In this appendix, we prove these equations and briefly inspect the profile of the likelihood function.
The effect of the Gaussian process is also described by visualizing the distributions of the likelihood and posterior probability of Bayesian estimation.
        
The exponent in equation (\ref{eq:likelihood}) is transformed as
\begin{equation}
    \begin{split}
        -\frac{1}{2} \sum_l \frac{\qty{\eta_l^{TA} - \qty(\boldsymbol{C} \boldsymbol{\theta})_l}^2}{\qty(\sigma_l^{TA})^2}
        &= -\frac{1}{2} \sum_l \frac{1}{\qty(\sigma_l^{TA})^2} \qty{
           \sum_{k=1}^{N}\sum_{k'=1}^{N} C_{lk} C_{lk'} \theta_k \theta_{k'}
           - 2 \sum_{k=1}^{N} \eta_l^{TA} C_{lk} \theta_k
           + \qty(\eta_l^{TA})^2}\\
        &= -\frac{1}{2} \qty{ \sum_{k=1}^{N}\sum_{k'=1}^{N} \theta_k
           \qty(\sum_l \frac{C_{lk} C_{lk'}}{\qty(\sigma_l^{TA})^2}) \theta_{k'}
           - 2 \sum_{k=1}^{N} \theta_k \sum_l \frac{C_{lk} \eta_l^{TA}}{\qty(\sigma_l^{TA})^2}
           + \sum_l \qty(\frac{\eta_l^{TA}}{\sigma_l^{TA}})^2}\\
        &= -\frac{1}{2} \qty( \boldsymbol{\theta}^\top \boldsymbol{Q} \boldsymbol{\theta}
           - 2 \boldsymbol{\theta}^\top \boldsymbol{p} + u )
    \end{split}
    \label{eq:likelihood-exp}
\end{equation}
where $u = \sum_l \qty(\eta_l^{TA} / \sigma_l^{TA})^2$ and $\boldsymbol{Q}$ and $\boldsymbol{p}$ are defined in equation (\ref{eq:likelihood-param}).
On the other hand, a square-completed quadratic form for $\boldsymbol{\theta}$ is defined and expanded as
\begin{equation}
    \begin{split}
        \qty(\boldsymbol{\theta} - \boldsymbol{\Theta}_L)^\top \boldsymbol{Q} \qty(\boldsymbol{\theta} - \boldsymbol{\Theta}_L) + u'
        &= \boldsymbol{\theta}^\top \boldsymbol{Q} \boldsymbol{\theta}
            - \boldsymbol{\theta}^\top \boldsymbol{Q} \boldsymbol{\Theta}_L
            - \boldsymbol{\Theta}_L^\top \boldsymbol{Q} \boldsymbol{\theta}
            + \boldsymbol{\Theta}_L^\top \boldsymbol{Q} \boldsymbol{\Theta}_L + u'\\
        &= \boldsymbol{\theta}^\top \boldsymbol{Q} \boldsymbol{\theta}
            - \boldsymbol{\theta}^\top
            \qty( \boldsymbol{Q} \boldsymbol{\Theta}_L + \boldsymbol{Q}^\top \boldsymbol{\Theta}_L )
            + \boldsymbol{\Theta}_L^\top \boldsymbol{Q} \boldsymbol{\Theta}_L + u'\\
        &= \boldsymbol{\theta}^\top \boldsymbol{Q} \boldsymbol{\theta}
            - 2 \boldsymbol{\theta}^\top \boldsymbol{Q} \boldsymbol{\Theta}_L
            + \boldsymbol{\Theta}_L^\top \boldsymbol{Q} \boldsymbol{\Theta}_L + u'
    \end{split}
    \label{eq:likelihood-square}
\end{equation}
where $\boldsymbol{Q}^\top = \boldsymbol{Q}$ is used.
Comparing the parentheses in equation (\ref{eq:likelihood-exp}) and equation (\ref{eq:likelihood-square}) for $\boldsymbol{\theta}$, we obtain $\boldsymbol{p} = \boldsymbol{Q}\boldsymbol{\Theta}_L$, which is also identical to the definition of $\boldsymbol{\Theta}_L$ in equation (\ref{eq:likelihood-param}).
The parameter $u'$ is also defined from the comparison as $u' = u - \boldsymbol{\Theta}_L^\top \boldsymbol{Q} \boldsymbol{\Theta}_L$.
Replacing the parentheses in equation (\ref{eq:likelihood-exp}) with the square-completed expression in equation (\ref{eq:likelihood-square}) and revisiting the conditional probability distribution $\mathcal{P}\qty(\boldsymbol{\eta} | \boldsymbol{\theta})$ in equation (\ref{eq:likelihood}), we obtain
\begin{equation}
    \mathcal{L}\qty(\boldsymbol{\theta}) \propto \exp{- \frac{1}{2}
    \qty(\boldsymbol{\theta} - \boldsymbol{\Theta}_L)^\top \boldsymbol{\Sigma}_L^{-1} \qty(\boldsymbol{\theta} - \boldsymbol{\Theta}_L) + u' }
    \propto \mathcal{N}\qty(\boldsymbol{\theta} \;|\; \boldsymbol{\Theta}_L, \boldsymbol{\Sigma}_L)
\end{equation}
where $\mathcal{P}\qty(\boldsymbol{\eta} | \boldsymbol{\theta})$ is replaced by a symbol $\mathcal{L}\qty(\boldsymbol{\theta})$ to express the likelihood function for $\boldsymbol{\theta}$.
The definition $\boldsymbol{\Sigma}_L = \boldsymbol{Q}^{-1}$ in equation (\ref{eq:likelihood-param}) is also used.
Therefore, it is proven that the likelihood function is proportional to a multivariate normal distribution for $\boldsymbol{\theta}$ where parameters $\boldsymbol{\Theta}_L$ and $\boldsymbol{\Sigma}_L$ in equation (\ref{eq:likelihood-param}) are identical to the mean vector and covariance matrix of the distribution, respectively.
The posterior distribution in equation (\ref{eq:bayes}) is equivalent to a product of multivariate normal distributions $\mathcal{L}\qty(\boldsymbol{\theta}) \propto \mathcal{N}\qty(\boldsymbol{\theta} \;|\; \boldsymbol{\Theta}_L, \boldsymbol{\Sigma}_L)$ and $\mathcal{P}\qty(\boldsymbol{\theta}) \propto \mathcal{N}\qty(\boldsymbol{\theta} \;|\; \boldsymbol{\Theta}_G, \boldsymbol{\Sigma}_G)$.
From a formula for a product of multivariate normal distributions, the mean vector $\boldsymbol{\Theta}$ and covariance matrix $\boldsymbol{\Sigma}$ of the posterior distribution are derived from $\boldsymbol{\Theta}_L$, $\boldsymbol{\Sigma}_L$, $\boldsymbol{\Theta}_G$, and $\boldsymbol{\Sigma}_G$, as described in equation (\ref{eq:posterior-param}).

\begin{figure}
    \centering
    \includegraphics[width=0.6\textwidth]{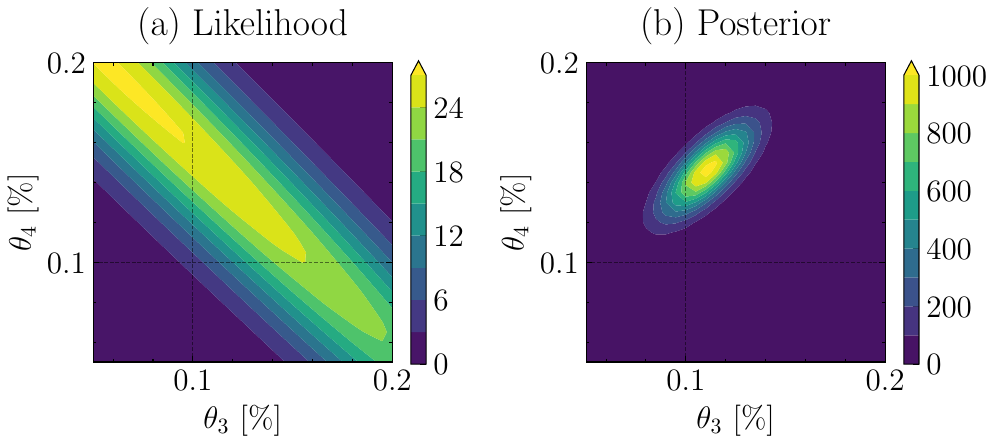}
    \caption{Conditional distributions of the (a) likelihood of observed data and (b) posterior probability of Bayesian estimation, each in $\theta_3$ - $\theta_4$ space for the sample year 2015.
    Other parameters, $\theta_k$ for $k \ne 3 \;\rm{or}\; 4$, are fixed at those in the mean vector $\boldsymbol{\Theta}$ of the posterior distribution.
    Each distribution is normalized so that its integration in the $\theta_3$ - $\theta_4$ space is equal to 1.}
    \label{fig:likelihood}
\end{figure}
Figure \ref{fig:likelihood}a shows a conditional distribution of the likelihood $\mathcal{L}\qty(\boldsymbol{\theta})$ in the $\theta_3$ - $\theta_4$ parameter space for the sample year 2015, where other parameters, $\theta_k$ for $k \ne 3 \;\rm{or}\; 4$, are fixed at those in the mean spectrum $\boldsymbol{\Theta}$ derived by equation (\ref{eq:posterior-param}).
The distribution is normalized so that its integration in the $\theta_3$ - $\theta_4$ space equals 1.
The likelihood distribution is widely extended along a $\theta_3 + \theta_4 = const.$ line, indicating that these parameters cancel out each other in the likelihood.
This demonstrates the difficulty to uniquely determine these parameters, or the spectrum values in adjacent rigidity bins by using the likelihood function.
Figure \ref{fig:likelihood}b demonstrates how the Gaussian process works to overcome this problem.
It is a $\theta_3$ - $\theta_4$ parameter space distribution of the posterior probability $\mathcal{P} \qty( \boldsymbol{\theta} | \boldsymbol{\eta} )$ in 2015, derived by multiplying the Gaussian process $\mathcal{P}\qty(\boldsymbol{\theta})$ to the conditional probability $\mathcal{P} \qty( \boldsymbol{\eta} | \boldsymbol{\theta} )$, or likelihood function $\mathcal{L}\qty(\boldsymbol{\theta})$, in equation (\ref{eq:bayes}). 
The high-confidence region is shrunk in Figure \ref{fig:likelihood}b compared to Figure \ref{fig:likelihood}a, within a reasonable region where the adjacent rigidity bins $\theta_3$ and $\theta_4$ have comparable values with each other.
This limitation by the Gaussian process prevents a discontinuous spectrum, enabling us to determine the spectrum while keeping a sufficient agreement between the derived spectrum and observed data.
On the other hand, how the likelihood works is visualized by the more significant likelihood in the $\theta_4>\theta_3$ region than $\theta_4<\theta_3$ region in Figure \ref{fig:likelihood}a.
It suggests that the spectrum value $\theta_k$ increases with the rigidity or its index $k$, resulting in the hard spectrum in 2015 as displayed in Figure \ref{fig:sample_years}b.

\cite{fujimoto1983} also approximated a rigidity spectrum of the semi-diurnal anisotropy amplitude by a step function, as well as our formulation for the NS anisotropy in equation (\ref{eq:step}).
Then, they derived the spectrum by a least-square method which is equivalent to the maximum likelihood estimation.
In their analysis, the spectrum was split into only seven rigidity bins in a range from $\sim10$ GV to $\sim2000$ GV, and not only ground-based muon detectors but also underground detectors were used.
This rough rigidity resolution and expanded dataset likely allowed them to derive the spectrum only by the likelihood function $\mathcal{L}\qty(\boldsymbol{\theta})$ without the Gaussian process used in this study.
However, this approach is a trade-off with some accuracy, such as the rigidity bin and temporal resolution.
They derive only a several-year average of the spectrum, unlike our analysis deriving a solar cycle variation of the NS anisotropy spectrum on a yearly basis.

\section{Hyperparameters of the Gaussian process}\label{sec:hyper}
In this appendix, we visualize hyperparameter dependences of the Gaussian process and describe a hyperparameter tuning performed for this study.
As described in Section \ref{subsec:bayes}, the Gaussian process expresses the probability density function of the rigidity spectrum $\boldsymbol{\theta}$ by a multivariate normal distribution $\mathcal{N} \qty(\boldsymbol{\theta} \;|\; \boldsymbol{\Theta}_G, \boldsymbol{\Sigma}_G)$.
The covariance matrix $\boldsymbol{\Sigma}_G$ defines the correlation coefficients of the spectrum values across individual rigidity bins, confining the acceptable range of the smoothness of the spectrum.
The correlation coefficient is determined by the hyperparameter $b$ in equation (\ref{eq:cov_gauss}).
The coefficient is reduced along with the rigidity difference between rigidity bins $k$ and $k'$, $|q_k - q_{k'}|$ in equation (\ref{eq:cov_gauss}), and reaches $1/e \sim 0.37$ where $|q_k - q_{k'}| = b$ [$\log_{10}({\rm GV})$].

\begin{figure}
    \centering
    \includegraphics[width=0.8\textwidth]{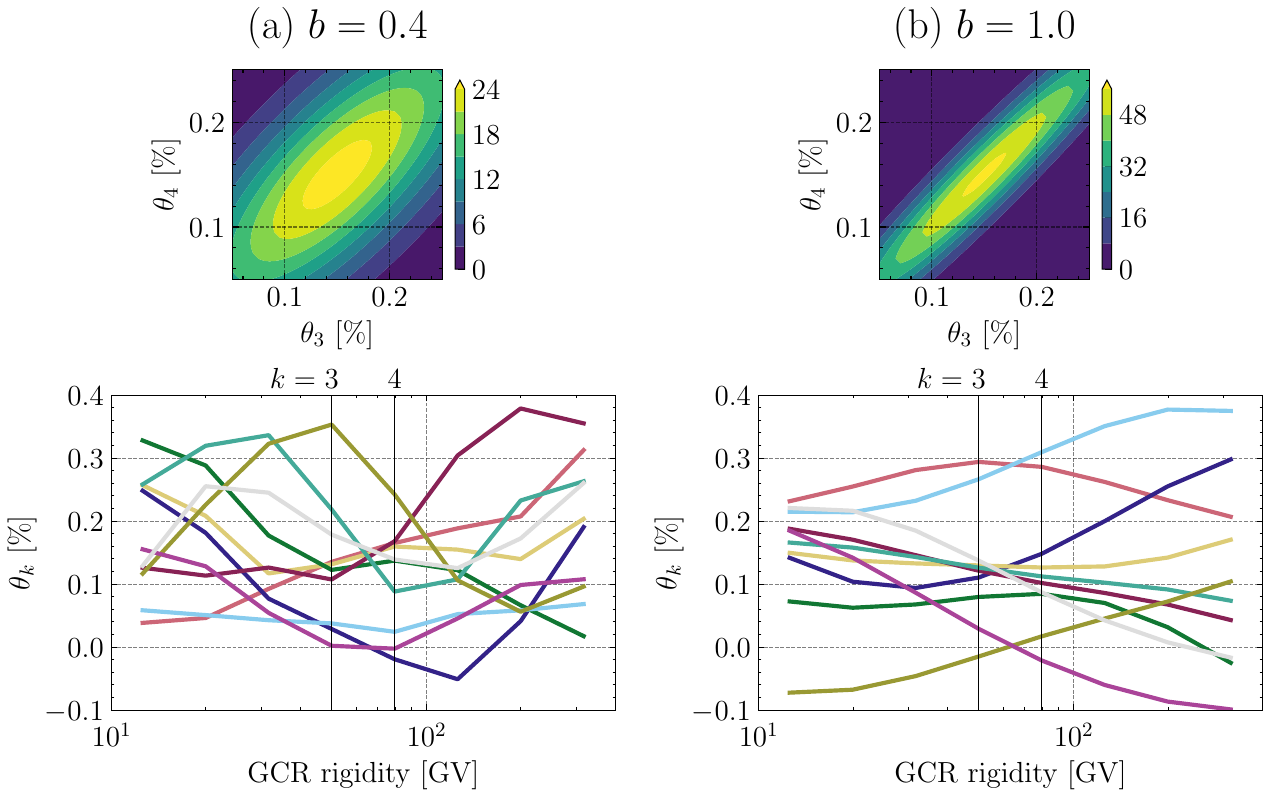}
    \caption{Probability density distribution projected onto the $\theta_3$ - $\theta_4$ parameter space (upper panel) and 10 random spectra (lower panel) for the Gaussian process with a hyperparameter (a) $b=0.4$ or (b) $b=1.0$ [$\log_{10}({\rm GV})$], respectively. Other hyperparameters are set at common values, $\theta^c=0.15\%$ and $\sigma_G=0.1\%$, for both cases (a) and (b).}
    \label{fig:gauss}
\end{figure}
The upper panel in Figure \ref{fig:gauss}a displays the $\theta_3$ - $\theta_4$ parameter space distribution of the Gaussian process $\mathcal{N} \qty(\boldsymbol{\theta} \;|\; \boldsymbol{\Theta}_G, \boldsymbol{\Sigma}_G)$, where the hyperparameters $\sigma_G$ and $b$ in equation (\ref{eq:cov_gauss}) are set at those used in Section \ref{subsec:bayes}, $\sigma_G=0.1\%$ and $b=0.4$ [$\log_{10}({\rm GV})$].
Components of the mean vector $\boldsymbol{\Theta}_G$ are set at a constant value $\theta^c = 0.15\%$.
From Table \ref{tab:bins}, the rigidity difference between rigidity bins $k=3$ and $k'=4$ is $|q_k - q_k'| = 0.2$ [$\log_{10}({\rm GV})$].
This results in the correlation coefficient $\exp \qty(-|q_k - q_k'|^2/b^2) = 0.78$, visualized by the probability density distribution along the $x=y$ line in the upper panel of Figure \ref{fig:gauss}a.
The lower panel in Figure \ref{fig:gauss}a displays 10 random spectra $\boldsymbol{\theta}$'s following the probability distribution of the Gaussian process.
Each vertical solid line denotes the median rigidity $P_k^m$ [GV] in the rigidity bin $k=3$ or 4.
An average level of the spectra is determined by $\boldsymbol{\Theta}_G$ or $\theta^c$ at 0.15\%, while the deviation of spectrum values is limited from $\sim0.0\%$ to $\sim0.3\%$ by the hyperparameter $\sigma_G$.
In each line, discontinuity of the spectrum values in adjacent rigidity bins, such as $\theta_3$ and $\theta_4$, is confined by the correlation coefficient determined by $b$.

Figure \ref{fig:gauss}b is the same as Figure \ref{fig:gauss}a but only the hyperparameter $b$ is set at a different value $b=1.0$ [$\log_{10}({\rm GV})$].
The correlation coefficients between spectrum values across rigidity bins become more significant than the case of $b=0.4$ in Figure \ref{fig:gauss}a, where $\exp \qty(-|q_k - q_k'|^2/b^2) = 0.96$ between $\theta_3$ and $\theta_4$.
This high coefficient leads to smoother rigidity spectra than the case of $b=0.4$, or nearly linear spectra, as demonstrated by the lower panel of Figure \ref{fig:gauss}b. 
In this manner, we can confine the smoothness of the spectrum by introducing the Gaussian process as a prior distribution in Bayesian estimation as performed in Section \ref{subsec:bayes}, without assuming any analytical function.

Hyperparameters of the Gaussian process have to be predefined when we perform Bayesian estimation of the rigidity spectrum in equation (\ref{eq:bayes}).
The mean vector $\boldsymbol{\Theta}_G$ is determined every year as follows.
In the same way as Appendix \ref{sec:sense} and equation (\ref{eq:eta-flat}), assuming a flat spectrum, $\theta_1 \sim \theta_2 \sim \ldots \sim \theta_N \sim \theta^c = const.$, simplifies equation (\ref{eq:exp}) as 
\begin{equation}
    \tilde\eta_l^{TA} \sim c_l^z \theta^c
\end{equation}
where $c_l^z = \sum_{k=1}^N C_{lk}$.
In this case, the NS anisotropy $\theta^c$ is estimated as a weighted mean of the observed value $\eta_l^{TA}$ divided by $c_l^z$, as
\begin{equation}
    \theta^c \sim \frac{\sum_l w_l \eta_l^{TA} / c_l^z}{\sum_l w_l}
    \label{eq:prior_mean}
\end{equation}
where $w_l = \qty(c_l^z / \sigma_l^{TA})^2$.
We adopt $\Theta_{G,1} = \Theta_{G,2} = \ldots = \Theta_{G,N} = \theta^c$ for $\boldsymbol{\Theta}_G$ using the observed value $\eta_l^{TA}$ in each year, as a reference to the average level of the spectrum.

\begin{figure}
    \centering
    \epsscale{0.6}
    \plotone{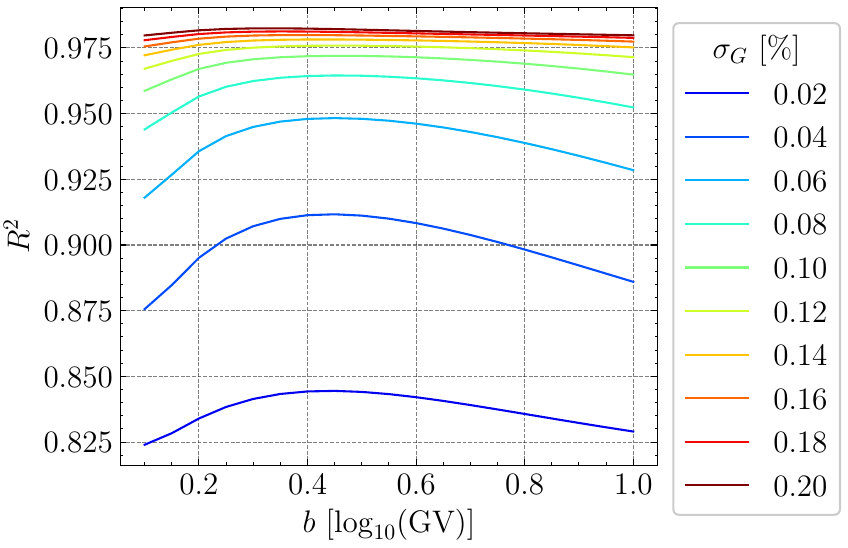}
    \caption{Result of the spectrum estimation for various combinations of the hyperparameters, $b$ [$\log_{10}$(GV)] and $\sigma_G$ [\%], in the Gaussian process. The coefficient of determination, $R^2$, of the mean spectrum $\Theta$ for the observed values $\eta_l^{TA}$'s in the sample year 2015 is displayed as a function of $b$ for each $\sigma_G$.}
    \label{fig:tuning}
\end{figure}
On the other hand, we use common values of $\sigma_G$ and $b$ for all years to ensure consistency of the precondition in this study. 
Too loose constraints of the Gaussian process, corresponding to larger $\sigma_G$ and smaller $b$, cause an over-fitting of the spectrum to observed data.
On the other hand, too strict constraints cause an under-fitting, and the reproducibility of the observed values by the derived spectrum is degraded.
From the mean spectrum $\boldsymbol{\Theta}$ of the posterior distribution derived in equation (\ref{eq:posterior-param}), a coefficient of determination is defined as
\begin{equation}
    R^2 = 1 - \frac{\sum_l \qty(\eta_l^{TA} - \tilde\eta_l^{TA})^2}{\sum_l \qty(\eta_l^{TA} - \left<\eta_l^{TA}\right>)^2}
\end{equation}
where $\tilde\eta_l^{TA}$ is an expected value of the observable $\eta_l^{TA}$ derived by inserting the mean spectrum $\boldsymbol{\Theta}$ as $\boldsymbol{\theta}$ in equation (\ref{eq:exp}).
An average of $\eta_l^{TA}$ for all channel pairs $l$'s is expressed by $\left<\eta_l^{TA}\right>$.
Now we focus on observed data $\eta_l^{TA}$'s in 2015 as a sample year.
For this year's data, the posterior distribution is calculated by equation (\ref{eq:posterior-param}) for all combinations of $\sigma_G$ and $b$ in a sufficiently wide range of these parameters.
Then, we derived $R^2$ for each combination of the hyperparameters.
Figure \ref{fig:tuning} displays the result, $R^2$, as a function of $b$ for each $\sigma_G$ value.
An increase of $R^2$ with increasing $\sigma_G$ saturates around $\sigma_G \sim 0.1\%$, indicating that a constraint as strict as $\sigma_G \sim 0.1\%$ can be imposed without losing $R^2$.
In overall lines in Figure \ref{fig:tuning} for $\sigma_G \le 0.1\%$, $R^2$ is maximized around $b = 0.4$ [$\log_{10}({\rm GV})$].
Based on these inspections, we adopt $\sigma_G=0.1\%$ and $b = 0.4$ [$\log_{10}({\rm GV})$] as optimal hyperparameters of the Gaussian process for our dataset.

\bibliography{main}{}
\bibliographystyle{aasjournal}

\end{document}